\documentclass[aps,prb,onecolumn,notitlepage,amsmath,amstex,amssymb,longbibliography,floatfix,a4paper]{revtex4-1}

\usepackage{hyperref}
\hypersetup{
	colorlinks=true,
	linkcolor=blue,
	filecolor=magenta,      
	urlcolor=cyan,
	pdftitle={Weak ergodocity breaking through the lens of quantum entanglement}
}

\usepackage{graphicx}
\usepackage{framed}
\usepackage{color,times}
\usepackage{xcolor}
\usepackage{dsfont}
\usepackage{afterpage}	
\usepackage{braket}
\usepackage{mathtools}
\usepackage{enumerate}
\usepackage{comment}

\definecolor{darkblue}{rgb}{0, 0, 0.8}

\begin{document}

\tolerance 10000

\title{Weak ergodicity breaking through the lens of quantum entanglement}
\author{Zlatko Papi\'c}
\affiliation{School of Physics and Astronomy, University of Leeds, Leeds LS2 9JT, United Kingdom}
\date{\today}

\begin{abstract}
Recent studies of interacting systems of quantum spins, ultracold atoms and correlated fermions have shed a new light on how  isolated many-body systems can avoid rapid equilibration to their thermal state. It has been shown that many such systems can ``weakly" break ergodicity:  they possess a small number of non-thermalising eigenstates and/or display slow relaxation from certain initial conditions, while majority of other initial states equilibrate fast, like in conventional thermalising systems. In this chapter, we provide a pedagogical introduction to weak ergodicity breaking phenomena, including Hilbert space fragmentation and quantum many-body scars. Central to these developments have been the tools based on quantum entanglement, in particular matrix product states and tangent space techniques, which have allowed to analytically construct non-thermal eigenstates in various non-integrable quantum models, and to explore semiclassical quantisation of such systems in the absence of a large-$N$ or mean field limit.  We also discuss recent experimental realisations of weak ergodicity breaking phenomena in systems of Rydberg atoms and tilted optical lattices.
\end{abstract}

\maketitle

\section{Introduction}\label{sec:intro}

\begin{figure*}[tb]
	\includegraphics[width=0.95\textwidth]{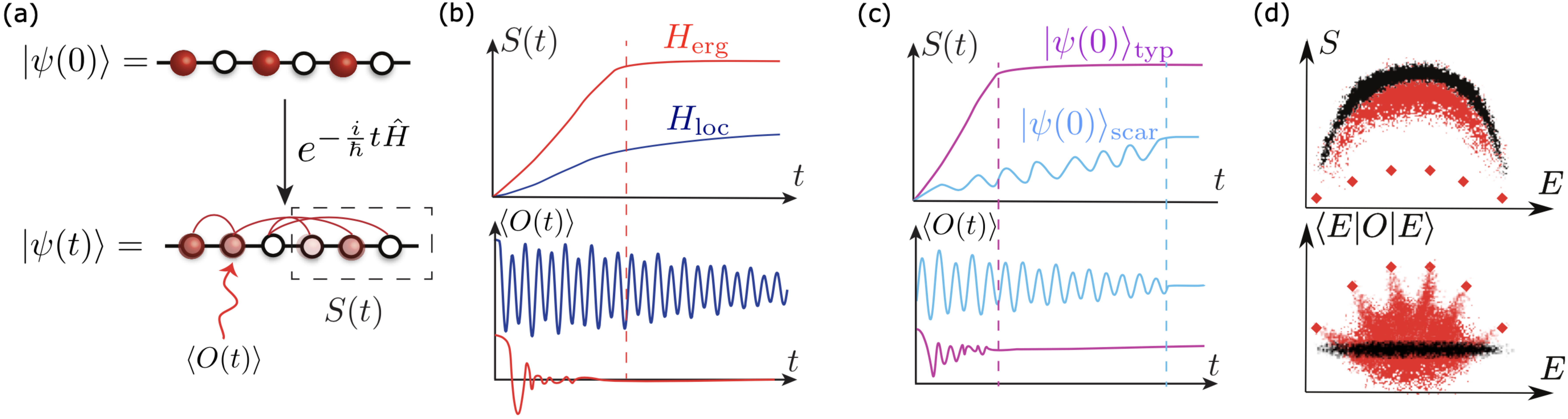}\\
	\caption{ Strong vs. weak breakdown of thermalisation.
		(a) Experiments probe thermalisation of isolated many-body systems using a quantum quench: the system is prepared in a simple initial state $|\psi(0)\rangle$ and the dynamics of some local observable $O$ and entanglement entropy $S$ are measured during unitary evolution. 
		(b) Strong breakdown of ergodicity caused by (many-body) localisation leads to qualitatively different dynamical behaviour compared to an ergodic system. This difference persists for broad classes of  initial states.
		(c) In contrast to (b), \emph{weak} ergodicity breaking results in strikingly diffferent dynamical behaviour when  \emph{different} initial states evolve under the same thermalising Hamiltonian. Quantum scarred systems, discussed in Sec.~\ref{sec:scars}, are an important class of systems where such behaviour has been experimentally observed. 
		(d): Manifestations of weak ergodicity breaking  in entanglement and expectation values of local observables in the system's eigenstates. Red dots correspond to eigenstates of a weakly non-ergodic system, where enhanced fluctuations and outlier states (red diamonds) are visible, compared to typical eigenstates of conventional thermalising systems (black dots) whose properties vary smoothly with energy $E$. This chapter focuses on the mechanisms and physical realisations of the phenomenology summarised in (c)-(d).
	}
	\label{Fig1} 
\end{figure*}

Experimental progress in cold atoms, trapped ions and superconducting circuits~\cite{Bloch2012,quantumsimulationRMP14} has generated a flurry of  interest into foundational questions of many-body quantum mechanics, such as how a statistical-mechanics description emerges in  isolated quantum systems comprising many degrees of freedom. In experiments with quantum simulators, the process of thermalisation can be conveniently probed in real time by quenching the system: one prepares a non-equilibrium initial state $|\psi(0)\rangle$, typically a product state of atoms, and observes its fate after time $t$, see Fig.~\ref{Fig1}(a). 

Well-isolated systems can be assumed to evolve according to the Schr\"odinger equation for the system's Hamiltonian~$H$.  Thermalisation can be characterised by the 
time evolution of local observables, $\langle O(t) \rangle$, or  entanglement entropy, $S(t) {=} {-}\mathrm{tr}\rho_A(t) \ln \rho_A(t)$. The  latter is defined as the von Neumann entropy for the reduced density matrix, $\rho_A$, of the subsystem $A$. Here we assume that the entire system is bipartitioned into $A$ and $B$ subsystems, and $\rho_A(t){=}\mathrm{tr}_B \ket{\psi(t)}\bra{\psi(t)}$ is obtained by tracing out the degrees of freedom belonging to $B$. Entanglement entropy quantifies the spreading of quantum correlations between spatial regions as the entire system remains in a pure state.

The results of measurements described above are schematically illustrated in Fig.~\ref{Fig1}(b), which contrasts the behaviour of two large classes of physical systems: quantum-ergodic systems~\cite{DeutschETH,SrednickiETH}  and many-body localised (MBL) systems~\cite{Mirlin05,Basko06,PalHuse}.  In the first case, parts of the system act as heat reservoirs for other parts, and an initial non-equilibrium state relaxes to a thermal equilibrium, with some well-defined effective temperature.  This behaviour is reminiscent of classical chaotic systems, which effectively ``forget" their initial condition in the course of time evolution. In contrast, in MBL systems local observables reach some stationary value which is non-thermal, retaining the memory of the initial state. As a result,  systems with strong ergodicity breaking can sustain much longer  quantum coherence. For example, the information stored in certain local observables would not decay in MBL systems, while it would rapidly decohere in an ergodic system [Fig.~\ref{Fig1}(b), bottom panel].   The behaviour of local observables is mirrored by that of quantum entanglement:  in thermalising systems, entanglement  typically grows linearly with time,  $S(t){\sim} t$, corresponding to quasiparticles moving at a finite speed~\cite{CalabreseQuench}. In MBL systems, quasiparticles are localised, with residual, exponentially-decaying interactions that lead to slow, logarithmic increase of entanglement, $S(t){\propto} \log t$~\cite{Znidaric08, Moore12, Serbyn13-2}, see the top panel of Fig.~\ref{Fig1}(b).

The generic behaviour of quantum-chaotic systems sketched in Fig.~\ref{Fig1}(b) is expected to hold irrespective of the chosen initial state. This is a consequence of  the Eigenstate Thermalisation Hypothesis (ETH)~\cite{DeutschETH,SrednickiETH},  a powerful conjecture governing the behaviour of quantum-ergodic systems -- see Box~1. In contrast, the present chapter focuses on recent theoretical works and experiments which pointed to the existence of a new class of behaviour, which can be loosely thought of as ``intermediate" between thermalisation in fully chaotic systems and strong ETH breakdown like in MBL systems. In such \emph{weakly} non-ergodic systems,  certain initial states show relaxation to thermal ensembles,  yet other states  exhibit non-stationary dynamics including persistent oscillations, illustrated in Fig.~\ref{Fig1}(c). Moreover, such systems also weakly violate the ETH in their eigenstate properties: e.g., they exhibit a few ``outlier" states with anomalously low entropy and local-observable matrix elements which do not vary smoothly with energy, see Fig.~\ref{Fig1}(d). 

This new regime of ergodicity breaking attracted broader attention after the experimental observation of dynamical revivals in large-scale Rydberg atom quantum simulators~\cite{Bernien2017}. Weak ergodicity breaking observed in such systems was subsequently named \emph{quantum many-body scarring}~\cite{Turner2017, wenwei18TDVPscar}, highlighting its analogy with chaotic stadium billiards, which had been known to host non-ergodic eigenfunctions bearing ``scars" of classical periodic orbits~\cite{Heller84}.  The discovery of many-body quantum scars fuelled a broader quest to understand physical consequences of weak ergodicity breaking. Importantly, during this quest, it came to  light that some examples of weak ergodicity breaking had previously been known. For example, non-integrable models such as the Affleck-Kennedy-Lieb-Tasaki (AKLT) spin chain, had been rigorously proven to possess non-thermal eigenstates~\cite{Bernevig2017,BernevigEnt}, which are now understood to share a similar algebraic structure with the non-thermal eigenstates in Rydberg atom systems. Other, more general, mechanisms of weak ergodicity breaking, such as Hilbert space fragmentation~\cite{Khemani2019_2,Sala2019} and ``embedding" constructions~\cite{ShiraishiMori} have also come into light and are being experimentally probed~\cite{Scherg2020}. 

This chapter provides a pedagogical introduction to weak ergodicity breaking phenomena, expanding upon a recent short overview in Ref.~\onlinecite{Serbyn2021}. As will become clear from the many physical examples presented below, the phenomenology of weak ergodicity breaking is quite rich and still rapidly evolving as this chapter is being written. Instead of giving an exhaustive review of all these developments, the aim is to highlight common themes between currently known examples of non-thermal dynamics, in particular focusing on the insights obtained by studying quantum entanglement in such systems. In Sec.~\ref{sec:mps} we start by introducing the relevant methodology based on matrix product states, that has successfully been used in recent works to analytically construct non-thermal eigenstates and to define the semiclassical limit of non-integrable many-body systems. In Sec.~\ref{sec:mechanisms} we discuss in detail three main mechanisms of weak ergodicity breaking and highlight their realisations in physical systems. Sec.~\ref{sec:pxp} introduces the so-called PXP model, which has played one of the key roles in understanding weak ergodicity breaking in Rydberg-atom experiments. In Sec.~\ref{sec:semiclassical} we relate the semiclassical quantisation of many-body systems with the  time-dependent variational principle applied to manifolds spanned by tensor network states. In Sec.~\ref{sec:scars} we explain how these ideas allow to explore parallels between weak ergodicity breaking in many-body systems and quantum scars in few-body systems. Finally, in Sec.~\ref{sec:exp} we discuss  two experimental platforms  -- Rydberg atoms and cold atoms in a tilted optical lattice -- that have recently observed signatures of weak ergodicity breaking. Conclusions and some open questions are presented in Sec.~\ref{sec:conclusions}.

\vspace*{0.5cm}
\definecolor{shadecolor}{rgb}{0.8,0.8,0.8}
\begin{shaded}
	\noindent{\bf Box 1 $|$ Eigenstate Thermalisation Hypothesis (ETH)}
\end{shaded}
\vspace{-9mm}
\definecolor{shadecolor}{rgb}{0.9,0.9,0.9}
\begin{shaded}
	\noindent Eigenstate thermalisation hypothesis (ETH)~\cite{DeutschETH,SrednickiETH} governs the process of thermalisation in closed quantum systems in the absence of coupling to a thermal bath. For present purposes, the following three consequences of the ETH are most relevant (for more information, see one of the recent reviews~\cite{Huse-rev,Gogolin2016,Mori2018}): 
	\begin{enumerate}
		\item The expectation values of physical observables in individual, highly-excited eigenstates of ergodic systems are ``thermal", i.e., they are identical to those evaluated using the microcanonical ensemble. Thus, 	highly-excited states of ergodic systems can be intuitively viewed as random vectors in the Hilbert space, and expectation values of local observables in such eigenstates are a smooth function of energy, independent of other microscopic details.
		\item By postulating the form of the off-diagonal matrix elements, the ETH makes predictions for the temporal fluctuations of local observables:  independent of the initial state, an observable approaches its equilibrium value, and then remains near that value most of the time, with fluctuations exponentially suppressed by the thermodynamic entropy. 
		\item  For a large finite subsystem $A$ of an infinite ETH system, the reduced density matrix $\rho_A$ is equal to the thermodynamic density matrix  at the effective temperature set by the energy of the corresponding eigenstate. 
	\end{enumerate}

The equality between thermal and reduced density matrices (3) implies that the entanglement entropy scales as the \emph{volume} of region $A$. For example, in one spatial dimension, volume law implies $S{\propto}L_A$. This reflects the fact that ergodic eigenstates are highly entangled, and agrees with the intuition that the ETH eigenstates are similar to random vectors. In some physical systems considered below, we will encounter global kinetic constraints, similar to those occuring in classical glasses~\cite{KCM-rev}. In such systems, the Hilbert space is globally constrained, and a  ``random" vector is understood to be one compatible with the constraints. Consequently, the entropy of such constrained random vectors can differ~\cite{Morampudi2020} from the so-called Page value attained by random qubit states~\cite{Page1993}.

		In systems which obey the so-called ``strong" ETH~\cite{Ueda2020}, the above properties are expected to hold for \emph{all} states~\cite{RigolNature}. On the other end, maximal violation of ETH is known to occur in integrable systems~\cite{Sutherland} and MBL systems~\cite{AbaninRev}, both of which have macroscopic numbers of conservation laws. Between these extremes lie the weak ergodicity breaking phenomena discussed in this chapter, where the majority of eigenstates follow the predictions of the ETH while a smaller number of eigenstates (e.g., polynomially many in system size) violate these properties. Similarly, out-of-equilibrium dynamics from most initial conditions results in fast relaxation, consistent with ETH expectations, while special initial conditions can lead to non-stationary dynamics at relatively long times.
	
\end{shaded}

\section{Matrix product state methods}\label{sec:mps}

The investigation of weak ergodicity breaking in non-integrable quantum systems has greatly benefited from matrix product states (MPS)~\cite{PerezGarcia}, a formalism designed to compactly represent and perform algebraic manipulations on a class of quantum states that are weakly entangled. These methods are naturally suited to capture aspects of weak ergodicity breaking, as the main signatures of the latter are the suppression of entanglement or its growth rate compared to conventional thermalising systems. Recent works have demonstrated the utility of MPS methods in two new settings: (i) the MPS have allowed to \emph{exactly} construct highly-excited eigenstates (i.e., eigenstates at finite energy density above the system's ground state) of a wide class of physical systems; (ii) the MPS have been used to effectively \emph{define} the system's semiclassical dynamics by projecting the Schro\"odinger time evolution into the manifold of MPS states whilst conserving the total energy of the system. In this section, we provide a brief overview of this methodology which yielded much of the physical insights into weak ergodicity breaking phenomena presented in subsequent sections.

\subsection{Towers of quasiparticles}\label{sec:GS}

Consider a one-dimensional quantum chain with a $d$-dimensional Hilbert space on each of the $L$ sites.  The many-body basis of the system is formed by tensor products of single-site Hilbert spaces, $\ket{\sigma_1, \sigma_2, \cdots, \sigma_L}$. Any state can be decomposed in this basis as
\begin{eqnarray}\label{eq:manybodystate}
|\psi\rangle = \sum_{\sigma_1,\sigma_2,\ldots,\sigma_L} c_{\sigma_1\sigma_2\ldots \sigma_L} \ket{\sigma_1,\sigma_2,\ldots,\sigma_L}
\end{eqnarray} 
by specifying its $d^L$ coefficients $c_{\sigma_1\sigma_2\ldots \sigma_L}$. This illustrates the ``exponential barrier" to studying many-body systems: one must specify $\sim \exp(L)$ coefficients to fully describe a generic state of a system of size $L$.  

The idea of MPS is that each of the coefficients in Eq.~(\ref{eq:manybodystate}) can be viewed as resulting from a product of matrices $A_i^{[\sigma_i]}$ of dimensions $\chi\times \chi$,
\begin{eqnarray}\label{eq:mpsobc}
	c_{\sigma_1\sigma_2\ldots \sigma_L} = b_l^T A_1^{[\sigma_1]} A_2^{[\sigma_2]} \ldots A_L^{[\sigma_L]} b_r.
\end{eqnarray}
The matrices $A$ are defined over the so-called auxiliary space with dimension $\chi$, but they also depend on the physical degrees of freedom $\sigma_i$. Moreover, for open boundary conditions, the matrices also vary from site to site in general, hence they carry a label $i$. The boundary $\chi$-dimensional vectors, $b_l$ and $b_r$, determine the boundary conditions for the wavefunction. Thus, we have reexpressed a $d^L$-dimensional tensor $c_{\sigma_1\sigma_2\ldots \sigma_L}$  in terms of  $d\times\chi\times\chi$ tensors $A_i$. 

If the system is translation-invariant, the matrices are the same on every site (or between different unit cells, more generally) and the boundary vectors are replaced with a trace over the auxiliary space,
\begin{eqnarray}\label{eq:mpspbc}
	c_{\sigma_1\sigma_2\ldots \sigma_L} = \mathrm{Tr} \left( A_1^{[\sigma_1]} A_2^{[\sigma_2]} \ldots A_L^{[\sigma_L]} \right).
\end{eqnarray}
Eqs.~(\ref{eq:mpsobc})-(\ref{eq:mpspbc}) furnish an MPS representation for the wave function in Eq.~(\ref{eq:manybodystate}).

Intuitively, $\chi$ controls the degree of entanglement in the wave function. When $\chi{=}1$, we have a mean-field description where degrees of freedom on different sites are independent of each other; increasing $\chi$ builds in quantum correlations between different sites. For a general state, rewriting the coefficients $c_{\sigma_1\sigma_2\ldots \sigma_L}$ in MPS form would require $\chi$ to be exponentially large in $L$. However, for states in one spatial dimension that obey the ``area law"~\cite{EisertAreaLaw}, i.e., whose entanglement entropy obeys $S \leq \mathrm{const}$, $\chi$ is also bounded by a constant. This results in a major simplification when describing such states using the MPS. Moreover, the same language also offers a complete algebraic framework to manipulate the MPS, for example we can efficiently add  two MPS states or calculate expectation values of local observables sandwiched between MPS states. The latter is achieved by expressing local operators in a similar representation known as ``matrix product operator" (MPO)~\cite{VerstraeteReview}. 

Ground states of many important condensed-matter systems are known to possess elegant MPS representations. A notable example is the Affleck-Kennedy-Lieb-Tasaki (AKLT)~\cite{AKLT} quantum spin-1 model:
\begin{equation}\label{eq:AKLTHam}
	H = \sum_{j=1}^L \left(\frac{1}{3} + \frac{1}{2} \vec{S}_j \cdot \vec{S}_{j+1} + \frac{1}{6} (\vec{S}_j \cdot \vec{S}_{j+1})^2 \right),
\end{equation}
where $\vec{S}$ are the spin-1 operators and we assume periodic boundary condition, i.e., $L+1 \equiv 1$. The ground state of this model has an \emph{exact} MPS representation, Eq.~(\ref{eq:mpspbc}), with
\begin{eqnarray}
	A^{[+]} = \sqrt{\frac{2}{3}}\sigma^+, \; A^{[0]} =  -\frac{1}{\sqrt{3}}\sigma^z, \; A^{[-]} =  -\sqrt{\frac{2}{3}}\sigma^-,
	\label{eq:AKLTMPS}
\end{eqnarray}
where the labels $+$, $0$, and $-$ denote the $S_z {=} +1$, $0$, and $-1$ spin-1 basis states, respectively, and $\sigma^{\pm}$, $\sigma^z$ are the \emph{spin-1/2} Pauli operators. Thus, for the AKLT state, the physical dimension $d{=} 3$ and the bond dimension is only $\chi {=} 2$.  This state and the model in Eq.~(\ref{eq:AKLTHam}) played an important role in establishing the existence of the Haldane gap in integer-spin chains~\cite{Affleck1988}. 

Beyond examples such as AKLT, where the MPS \emph{exactly} describe certain states, it has been established more generally that ground-state wave functions of gapped local Hamiltonians can be \textit{approximated} by an MPS with a small bond dimension $\chi$~\cite{Verstraete2006}.  This result can be intuitively understood from the fact that the ground states of gapped local Hamiltonians are necessarily weakly entangled~\cite{Hastings07} and hence well-represented by the MPS. However, this intuition does not immediately extend to highly excited states at a finite energy density above the ground state. As previously discussed in Sec.~\ref{sec:intro},  in a generic quantum many-body system, finite-energy-density states are expected to be governed by the ETH, hence their entanglement is expected to be high (i.e., scaling with the volume of the subsystem), so their MPS representation would not be efficient (as mentioned above, it would require $\chi \propto \exp(L)$).

\begin{figure}[t]
	\includegraphics*[width=0.9\linewidth]{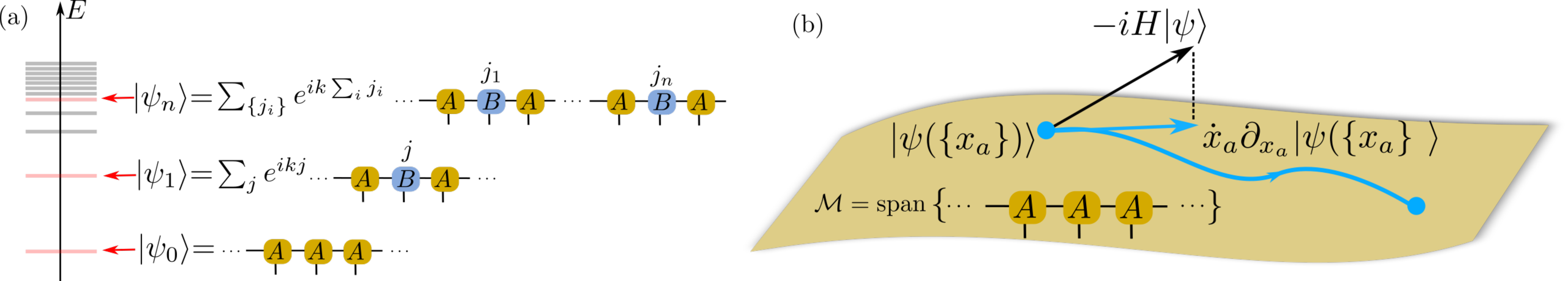}
	\caption{MPS methods for weak ergodicity breaking. (a) Constructions of exact, highly-excited eigenstates using MPO-MPS techniques from Ref.~\onlinecite{BernevigEnt}. This method allows to construct a tower of quasiparticle states $\ket{\psi_n}$ by acting with a local operator on the MPS ground state $\ket{\psi_0}$, as in Eq.~(\ref{eq:QPansatz}) and (\ref{eq:2qp}). (b) Time-dependent variational principle captures the optimal projection of quantum dynamics onto a given variational manifold $\mathcal{M}$, here parametrised by MPS. Limiting the bond dimension of the MPS to be low, this approach defines a semiclassical limit for quantum dynamics. The corresponding classical system is formed by equations of motions for the coefficients $x_a$, which parametrise the projection of the wave function into the MPS manifold. The equations of motions follow from minimising the leakage outside the manifold, i.e., the deviation of $-i H \ket{\psi(t)}$ from $\dot{x}_a\partial_{x_a} \ket{\psi(t)}$, see Eq.~(\ref{Eq:leak2}). }\label{fig:mps}
\end{figure}

Recent work by Moudgalya \emph{et al.}~\cite{BernevigEnt} on the AKLT model has shown that the MPS can nevertheless capture some ETH-violating eigenstates of a  non-integrable model, \emph{regardless} of how high in the energy spectrum such states occur. The technique used in these works in the construction of quasiparticle excitations above the ground state, originally introduced in works on tangent space methods~\cite{Haegeman2013, Vanderstraeten2015, Vanderstraeten2019}. Following Ref.~\onlinecite{BernevigEnt}, a single-site quasiparticle excitation with momentum $k$ on top of a general MPS state, $\ket{\psi_A}$, is given by
\begin{eqnarray}
	\ket{\psi_A\left(B, k\right)} = \sum_{j = 1}^{L} e^{i k j}\sum_{\{\sigma_j\}}     \textrm{Tr}\left(\cdots A^{[\sigma_{j-1}]} B^{[\sigma_j]} A^{[\sigma_{j+1}]} \cdots \right)  \ket{\sigma_1 \sigma_2 \cdots \sigma_L},
	\label{eq:QPansatz}
\end{eqnarray}
where $B^{[\sigma_j]}$ is a $\chi \times \chi$ matrix with physical dimension $d$ and $k$ denotes the momentum, see Fig.~\ref{fig:mps}(a).  In the framework of the Single-Mode Approximation~\cite{ArovasHaldaneAuerbach}, the quasiparticles are usually described in terms of a single-site ``quasiparticle creation operator" $\widehat{O}$, such that 
$	B^{[\sigma]} = \sum_{\sigma, \sigma'}{\widehat{O}_{\sigma, \sigma'} A^{[\sigma']}}$, which we denote in shorthand as $\ket{B} = \widehat{O}\ket{A}$. Note that this is a special case, as for example $\widehat{O}$ could act on several neighbouring sites -- such states have been considered in Ref.~\onlinecite{MoudgalyaFendley}.

To illustrate Eqs.~(\ref{eq:QPansatz}), some examples of low-lying excited eigenstates of the AKLT include the Arovas ``A" and ``B" states~\cite{Arovas1989}:
\begin{eqnarray}
	\ket{A} & =& \sum_{j=1}^L (-1)^j \vec{S}_j \cdot \vec{S}_{j+1} \ket{\psi_0},\\
	\ket{B} & = & \sum_{j=1}^L (-1)^j \{ \vec{S}_{j-1} \cdot \vec{S}_{j} , \vec{S}_j \cdot \vec{S}_{j+1} \}\ket{\psi_0},
\end{eqnarray}
where $\ket{\psi_0}$ is the AKLT ground state, $\{...\}$ denotes the anticommutator, and we have omitted the normalisation factors. Another exact excited state, as shown by Moudgalya \emph{et al.}~\cite{BernevigEnt}, is the spin-2 magnon state with momentum $\pi$, 
\begin{equation}\label{eq:akltexact}
	\ket{\psi_\mathrm{2-magnon}} = \sum_{j = 1}^{L}{(-1)^j (S^+_j)^2}\ket{\psi_0}. 
\end{equation}
All of these states can be written in the form of Eq.~(\ref{eq:QPansatz}). For example, the state in Eq.~(\ref{eq:akltexact}) can be expressed in such a form using the ground state matrices $A^{[\sigma]}$ in Eq.~(\ref{eq:AKLTMPS}) and the $B^{[\sigma]}$ matrices given by  $B^{[+]} = -\sqrt{\frac{2}{3}}\sigma^-$, $B^{[0]} = B^{[-]} = 0$. Note $B^{[+]}$ is the only non-trivial matrix, a direct consequence of the $(S^+)^2$ operator acting on spin-1.

In addition to single quasiparticles, multiple quasiparticle states can be described in the MPS formalism using multiple tensors.
For example, a state with two quasiparticles described by tensor $B$ with momentum $k$ is given by~\cite{BernevigEnt}
\begin{equation}\label{eq:2qp}
	\ket{\psi_A\left(B^2, k\right)} \equiv \left(\sum_{j}{e^{ikj} \widehat{O}_j}\right)^2\ket{\psi_0}. 
\end{equation}
Expanding the square, we see that we can have two cases: when $B$s act on the same site or different sites. Since in the AKLT chain we have $\widehat{O} = (S^+)^2$, we will have $\hat{O}^2\ket{\psi_0} = 0$ and the expression simplifies
\begin{eqnarray}
	\ket{\psi_A (B^2, k) } &=& \sum_{j_1\neq j_2} e^{i k (j_1+j_2)}     \sum_{\{m_j\}}    
 \textrm{Tr} \left( \cdots A^{[\sigma_{j_1-1}]} B^{[\sigma_{j_1}]} A^{[\sigma_{j_1+1}]} \cdots  
 \cdots   A^{[\sigma_{j_2-1}]} B^{[\sigma_{j_2}]} A^{[\sigma_{j_2+1}]} \cdots \right) \ket{\sigma_1 \sigma_2 \cdots \sigma_L} . \quad 
	\label{eq:QPansatz2}
\end{eqnarray}
Extending this formally to more quasiparticles appears straightforward. However, the difficulty that arises is that the successive applications of the quasiparticle operator lead to an increase in the bond dimension of the resulting state. Thus, one needs to finds a way to compress this state and compute its properties. These advances have been made in Ref.~\onlinecite{BernevigEnt}, allowing to evaluate the entanglement properties of highly excited states in the AKLT model, obtained by acting with $\hat O$ on the ground state an extensive number of times $n {\sim} L/2$.  The physical consequences of this for weak ergodicity breaking will be discussed further in Sec.~\ref{sec:sga} below.  

Finally, while the emphasis in this section has been on \emph{analytical} constructions of exact eigenstates containing finite density of quasiparticle excitations, one could envision employing similar methods to \emph{variationally} construct quasiparticle towers in other models which are less tractable than the AKLT. Indeed, such studies have been used to  numerically characterise dynamical properties of quantum spin systems such as the XXZ model, including quasiparticle dispersion relations and dynamical structure factor, albeit at much lower energies than the phenomena discussed here~\cite{VanderstraetenThesis}.

\subsection{Time-dependent variational principle}\label{sec:tdvp}

In the previous subsection, we mentioned that some non-integrable quantum spin chains have highly-excited energy eigenstates with a particularly simple structure, encoded in a local MPO acting on the ground state written as MPS. ``Simple" structure means that such eigenstates have sub-volume law entanglement, i.e., they are much less entangled than random vectors in the Hilbert space (cf. Box~1). Now we ask a complementary question: if the system undergoes anomalously slow unitary dynamics, e.g., the spreading of correlations is inhibited compared to a thermalising system, could we compactly describe such dynamics using the MPS? 
 
The natural framework to address the previous question is the time-dependent variational principle (TDVP), originally formulated by Dirac~\cite{Dirac1930}, which yields the optimal projection of quantum dynamics onto a given variational manifold, $\mathcal{M}=\mathrm{span}\{ |\psi(\mathbf{x})\rangle\}$, parametrized by some parameters $\mathbf{x} \equiv \{ x_a \} \in \mathbb{R}^N$, Fig.~\ref{fig:mps}(b). TDVP equations of motion are obtained by extremizing the action $\int dt\, \mathcal{L}$, with an effective Lagrangian~\cite{kramer1981geometry}
\begin{equation}\label{Eq:action}
	\mathcal{L}  =  \frac{i}{2}(\braket{\psi|\dot{\psi}}-\braket{\dot{\psi}|\psi})-\braket{\psi|H|\psi}.
\end{equation}
In the present case, we assume that $\mathcal{M}$ is spanned by the MPS of some given bond dimension, i.e., by matrices $A^\sigma(\mathbf{x})$ like in Eq.~(\ref{eq:mpspbc}) with bond dimension $\chi$ and the matrices depend on variational parameters, $\mathbf{x}{\in}\mathbb{R}^N$. These parameters can be given an interpretation in terms of pairs of coordinate and momenta (we assume $N$ is even). In Box~2, we motivate this approach using a very simple example of a single spin.

Extremising the action in Eq.~(\ref{Eq:action}) results in Euler-Lagrange equations of motion for the $\mathbf{x}$ variables~\cite{kramer1981geometry, Haegeman}:
\begin{equation}\label{Eq:TDVPMPS}
\sum_{a}\dot{{x}}_{a}\text{Im}\braket{\partial_{{x}_{b}}\psi|\partial_{{x}_{a}}\psi} = -\frac{1}{2}\partial_{{x}_{b}}\braket{\psi|H|\psi}.
\end{equation}
This set of equations constitutes an effective mapping of quantum dynamics onto a classical non-linear dynamical system.  

While TDVP has been successfully applied to manifolds spanned by the MPS~\cite{Haegeman} and finite tensor tree states~\cite{Bauernfeind2020}, in cases when it is possible to \emph{analytically} calculate  $\braket{\psi|\dot{\psi}}$ and $\braket{\psi|H|\psi}$, further insights can be obtained from studying the non-linear system using the tools of classical dynamical systems~\cite{StrogatzBook} and few-body chaos~\cite{GutzwillerBook}.   In practice, such studies are naturally limited to sufficiently simple manifolds which have low bond dimensions. Nevertheless, this approach proved extremely valuable in establishing an analogy between non-thermal eigenstates in Rydberg atom chains and their counterparts in few-body systems, justifying the name ``quantum many-body scars" -- these developments will be the subject of Secs.~\ref{sec:semiclassical} and \ref{sec:scars}.

We note that there are some caveats to mapping quantum dynamics onto low-dimensional manifolds $\mathcal{M}$. At late times,  the error of the TDVP approximation necessarily grows because the exact quantum dynamics generally brings the system out of the variational manifold -- see Fig.~\ref{fig:mps}(b).  Intuitively,  the TDVP error for a given initial state is linked to the rate of entanglement growth for that state, since the MPS with low bond dimension can only successfully capture weakly entangled states. When the entanglement grows significantly, the state starts to require an MPS description with a  bond dimension that extends beyond the variational manifold. This discrepancy is the error of the TDVP approximation, also known as \emph{ quantum leakage}~\cite{Haegeman,wenwei18TDVPscar}:
\begin{equation}\label{Eq:leak2}
\gamma^2(\{x_a\}) = \lim_{L\to\infty}\frac{1}{L}|| (iH + \dot{x}_b \partial_{x_b}) \ket{\psi(\{x_a\})} ||^2.
\end{equation}
This is the instantaneous rate at which the exact quantum wave function leaves the variational manifold. Equations of motion can be obtained by minimising the discrepancy between exact quantum dynamics and its projection onto $\mathcal{M}$. Explicit computation of $\gamma^2$ is possible within the TDVP framework, but since it involves the square of the Hamiltonian operator, it contains information that goes beyond the TDVP equations of motion. Understanding $\gamma^2$ is key to understanding the relation between quantum dynamics and its classical counterpart. We postpone  a more extensive discussion of the subtleties of the TDVP approach to many-body systems to Sec.~\ref{sec:semiclassical}.

\definecolor{shadecolor}{rgb}{0.8,0.8,0.8}
\begin{shaded}
	\noindent{\bf Box 2 $|$ TDVP for a single spin}
\end{shaded}
\vspace{-9mm}
\definecolor{shadecolor}{rgb}{0.9,0.9,0.9}
\begin{shaded}
	
	\noindent To illustrate the TDVP way of thinking, consider the simplest example of a quantum spin prepared in a product state $\ket{(\psi(\theta,\phi))}=\cos\theta \ket{\uparrow} + \sin\theta e^{i\phi} \ket{\downarrow}$, parametrised by angles $\theta$ and $\phi$ on the Bloch sphere. For concreteness, we can think of the spin Hamiltonian as $H{=}\epsilon \sigma^x$, corresponding to the magnetic field along the $x$-direction. Our aim is to find the equations of motion for the angles, $\theta(t)$ and $\phi(t)$, as the spin evolves under the influence of the magnetic field.
	
	The expectation value of the Hamiltonian in the given state is $\langle H \rangle = \epsilon \sin(2\theta)\cos\phi$. Moreover, we have $\langle \psi|\dot{\psi}\rangle = -i\dot{\phi}\sin^2\theta$. Plugging this into Eq.~(\ref{Eq:action}), our Lagrangian takes the form $\mathcal{L}=\dot{\phi}\sin^2\theta - \langle H \rangle$. We see that we can identify the canonical ``position" and ``momentum" variables as $\phi \to q$ and $\sin^2\theta \to p$. Thus, we have mapped the quantum spin dynamics onto a classical dynamical system described by the equations of motion 
	$\dot{\phi} = 2\epsilon\cot(2\theta)\cos\phi$ and $\dot{\theta} = -\epsilon \sin\phi$. Notice that the line $\phi{=}\pi/2$ is stationary -- this corresponds to the simple precession dynamics, $\dot{\theta}=-\epsilon$, which in this case is easy to derive directly by solving the Schr\"odinger equation. 
	
This example of a single spin can also be viewed as a particularly simple mean-field picture for the dynamics. This approach is  generally insufficient to describe dynamics in more complex many-body systems and, as discussed in Sec.~\ref{sec:semiclassical}, recent works have been exploring generalisations to variational manifolds beyond the mean-field using MPS states.

\end{shaded}

\section{Mechanisms of weak ergodicity breaking}\label{sec:mechanisms}

ETH-violating eigenstates have recently been theoretically identified in a variety of non-integrable quantum models, revealing a complex landscape of weak ergodicity breaking phenomena. The common feature of these models is the emergence of  a decoupled subspace within the many-body Hilbert space, in general without any underlying symmetry. In this section, we focus on idealised models where such a subspace is perfectly decoupled from the rest of the spectrum, resulting in a decomposition of the Hamiltonian 
\begin{equation}\label{Eq:decompose}
	H = H_{\mathrm{non-ETH}} \bigoplus H_{\mathrm{thermal}},
\end{equation}
where $H_\mathrm{non-ETH}$ is the non-thermalising subspace,  exactly decoupled from the thermalising bulk of the spectrum, $H_{\mathrm{thermal}}$. The eigenstates that inhabit the subspace $H_\mathrm{non-ETH}$ violate the ETH and have different properties compared to the majority of thermal eigenstates residing in $H_{\mathrm{thermal}}$.  Below we elucidate three commonly encountered  mechanisms that produce such decoupled subspaces. We would like to point out that in much of the recent literature, the non-thermalising eigenstates that span $H_{\mathrm{non-ETH}}$ are commonly referred to as ``quantum many-body scars". We will discuss the precise meaning of quantum many-body scarring in Sec.~\ref{sec:scars}.  

\subsection{Spectrum generating algebra}\label{sec:sga}

Using the MPS techniques introduced in Sec.~\ref{sec:mps}, Moudgalya \emph{at al.}~\cite{BernevigEnt}, building on the early work by Arovas~\cite{Arovas1989}, analytically constructed a tower of exact eigenstates in the AKLT model in Eq.~(\ref{eq:AKLTHam}), which were shown to have sub-volume entanglement entropy, thus providing the first rigorous demonstration of the strong-ETH violation. Later on, Schecter and Iadecola~\cite{Iadecola2019_2}  introduced a tower of exact eigenstates in a family of non-integrable spin-1 XY models as well in a a particular non-integrable spin-1/2 model that conserves the number of domain walls~\cite{Iadecola2019_3}.  In these and many other similar examples that followed, the non-thermal eigenstates form a so-called  ``spectrum generating algebra" (SGA), first introduced in the context of high energy physics~\cite{Arno1988} and subsequently applied to the Hubbard model~\cite{YangEta,ZhangEta}.

To define the SGA, we follow the presentation given by Mark, Lin and Motrunich~\cite{MotrunichTowers}. Suppose we have a Hamiltonian $H$; a linear subspace $W$; a state $\ket{\Psi_0} \in W$, which is an eigenstate of $H$ with energy $E_0$; and an operator $Q^\dagger$ such that $Q^\dagger W \subset W$ and
\begin{equation}
\left(\left[H, Q^\dagger \right] - \omega Q^\dagger \right) W = 0 ~.
\label{eq:sga}
\end{equation}
Then it can be easily proven that the family of states
\begin{equation}\label{eq:tower}
\ket{\mathcal{S}_n} = (Q^\dagger)^n \ket{\Psi_0},
\end{equation} 
as long as they are non-vanishing vectors, are eigenstates of $H$ with eigenvalues $E_0 {+} n \omega$. The non-thermalising nature of these states becomes non-trivial when  $W$ is \emph{not} the entire Hilbert space and $Q^\dagger$ is not associated with a symmetry of the Hamiltonian. The latter crucially distinguishes these non-thermal states  from somewhat similar ``eta-pairing" states in the Hubbard model~\cite{YangEta,ZhangEta}, a point discussed further at the end of this subsection.

As an example, in the AKLT model one of the possible SGA operators is  a spin-2 magnon excitation that we have previously seen in Eq.~(\ref{eq:akltexact}), with
\begin{eqnarray}\label{eq:aklttower}
	Q^\dagger = \sum_{j=1}^L (-1)^j (S^+_j)^2.
\end{eqnarray}
This generates the tower of states in Eq.~(\ref{eq:tower}) for $ n=0,...,L/2$, which extend from the ground state ($n{=}0$) up to the ferromagnetic state $\ket{1,1...,1}$ ($n{=}L/2$), and the state in Eq.~(\ref{eq:akltexact}) is a member of this family. It is also worth noting that these states are not unique, as they have total spin $s=2n$, hence we can obtain equivalent spin-rotated versions of these states by applying the SU(2) spin-lowering operator. 

Generally, it is clear that states of the form in Eq.~(\ref{eq:tower}) can be non-thermal, provided $|\Psi_0\rangle$ is sufficiently ``simple" (e.g., the ground state of $H$ if the latter is gapped) and $\hat Q^\dagger$ is a local operator. The AKLT tower of states in Eqs.~(\ref{eq:tower})-(\ref{eq:aklttower}) is the first example of a rigorous construction of a non-thermal family of states in a non-integrable model that are not protected by a global symmetry. Any Hamiltonian with the SGA property explicitly decomposes as in Eq.~(\ref{Eq:decompose}), where $H_\mathrm{non-ETH}$ contains the tower of states obtained by the action of $\hat Q^\dagger$.  Note that this algebraic structure is not powerful enough to fully diagonalise $H$; indeed, $H$ could well be a non-integrable Hamiltonian with an energy spectrum obeying the Wigner-Dyson statistics. 

Physically, $\hat Q^\dagger$ creates a wave packet corresponding to a ``quasiparticle" excitation
(e.g., a magnon), and repeated applications of $\hat Q^\dagger$ create a condensate of such quasiparticles.  In a class of frustration-free models which include  the AKLT~\cite{MoudgalyaFendley}, the quasiparticle condensates are non-thermal, e.g., their entanglement entropy scales logarithmically with the subsystem size:
\begin{eqnarray}\label{eq:logscaling}
	S \propto \log L_A.
\end{eqnarray}
 Another important clue when looking for non-thermalising SGA states is that they appear at energies which are integer multiples of $\omega$. Thus, SGA eigenstates can be detected as regularly spaced entropy-outliers, with entanglement much lower than the ETH volume law, as depicted in  Fig.~\ref{Fig1}(d).
Perhaps more practically, the SGA states also have very low entanglement \emph{rank}, i.e., many eigenvalues of $\rho_A$ strictly vanish, which is how these states were originally identified numerically in Ref.~\onlinecite{BernevigEnt}.

We mentioned  that the SGA traces back to the so-called ``eta-pairing states" in the Hubbard model~\cite{YangEta,ZhangEta}. For eta-pairing states, the subspace $W$ is the entire Hilbert space and $Q^\dagger$ essentially corresponds to a symmetry of the Hamiltonian~\cite{Vafek}. However, it is possible to perturb the Hubbard model~\cite{MarkHubbard,MoudgalyaHubbard} in a way which destroys the aforementioned symmetry and makes these states similar to the SGA states in the AKLT model discussed above. Note that the high degree of symmetry in Hubbard-type models can also lead to disconnected subspaces which host free-particle eigenstates with ballistic dynamics~\cite{ IadecolaZnidaric}. This ``fragmentation" of the Hilbert space is a much more general phenomenon beyond just the Hubbard model, as we discuss in the following section.
Finally, we note that  generalisations of the SGA construction can be found in a number of models in various dimensions~\cite{Iadecola2019_3, OnsagerScars, MotrunichTowers, Chattopadhyay,Lee2020,Pakrouski2020, Ren2020,Dea2020} (see Box~3).
It has also been pointed out that the SGA can arise in open quantum systems in the presence of dissipation or driving~\cite{Buca2019,BucaHubbard}.

\definecolor{shadecolor}{rgb}{0.8,0.8,0.8}
\begin{shaded}
	\noindent{\bf Box 3 $|$ Constructions of non-thermal towers of eigenstates}
\end{shaded}
\vspace{-9mm}
\definecolor{shadecolor}{rgb}{0.9,0.9,0.9}
\begin{shaded}

\noindent Towers of non-thermal eigenstates can be systematically constructed using the ``tunnels-to-towers" approach in Ref.~\onlinecite{Dea2020} based on generators of the Lie algebra of a symmetry group $G$.  For simplicity, consider the case when $G$ is SU(2) and we have a model defined by an SU(2)-symmetric Hamiltonian, $H_\mathrm{sym}$. The generators of the symmetry  $\{Q^+, Q^-, Q^z\}$ are associated with the corresponding su(2) algebra.  The spectrum of $H_\mathrm{sym}$ is organised into `tunnels' of degenerate eigenstates, with the same eigenvalue for the Casimir $Q^2$ but different eigenvalues for $Q^z$. One can `move' between states in a tunnel using $Q^\pm$.  

Now imagine perturbing the model by adding $H_\mathrm{SGA} \propto Q^z$.  This perturbation preserves the eigenstates, but breaks the degeneracy of the tunnels. Instead, states in each tunnel get promoted to `towers' and acquire an evenly spaced harmonic spectrum because of the SGA, $[Q^z, Q^\pm] = \pm Q^\pm$. Finally, we can further add a symmetry-breaking term $H_\mathrm{SB}$, such that it annihilates a specific tower of states but generically breaks all symmetries and mixes between the other states so as to make the rest of the spectrum thermal. In the full model, $H=H_\mathrm{sym} + H_\mathrm{SGA} + H_\mathrm{SB}$, our chosen tower of states are a collection of non-thermalising eigenstates, evenly distributed throughout the spectrum but not protected by a global symmetry, hence we arrive at a similar phenomenology to the AKLT model discussed in the main text. Various extension of this construction to non-Abelian groups and their $q$-deformed versions are possible~\cite{Dea2020}.

\vspace*{0.2cm}
\begin{minipage}[c]{0.98\linewidth}
	\begin{center}
		\includegraphics*[width=0.5\linewidth]{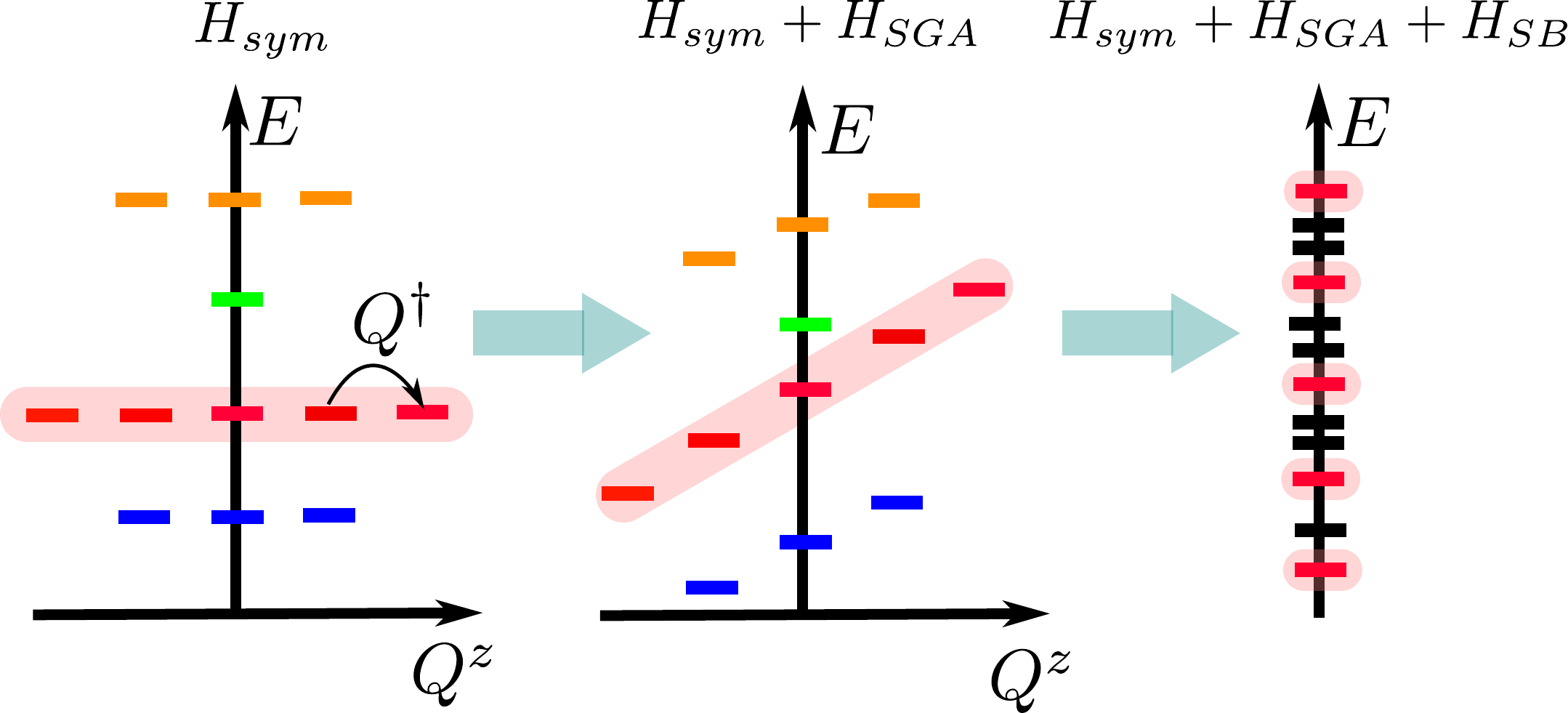}
	\end{center}
\begin{center}
	{\small {\bf Figure B3:} Tunnels-to-towers scheme for constructing non-thermal eigenstates from Ref.~\onlinecite{Dea2020}. }
\end{center}

\end{minipage}
\vspace*{0.2cm}

To illustrate the tunnels-to-towers construction, consider the spin-$1$ XY-like model introduced in Ref.~\onlinecite{Iadecola2019_2}:
\begin{eqnarray}
	H = \sum_{i} J(S^x_i S^x_{i+1} + S^y_i S^y_{i+1}) + J_3(S^x_i S^x_{i+3} + S^y_i S^y_{i+3}) 
	+ \sum_i h S^z_i +  D (S^z_i)^2.
	\label{Eq:1XY}
\end{eqnarray}	
 A tower of non-thermalising states is built by the action of the same $Q^\dagger$ as in the AKLT model~\cite{MotrunichTowers}:
\begin{equation}
	\ket{\mathcal{S}^{XY}_n} = \left(Q^+ \right)^n \ket{\Omega}, \quad Q^+ = \sum_j (-1)^j  \left(S^+_j \right)^2 ~,
	\label{eq:ISXYscar1}
\end{equation}
where $\ket{\Omega}$ is the fully polarised down state $|\Omega \rangle = |---\cdots-\rangle$. Note that $Q^+$ (and the corresponding $Q^-$), together with $Q^z = \frac{1}{2} [ Q^+, Q^- ]$, form an su(2) algebra but they are distinct from the spin-su(2) operators.  The first term $\propto J$ breaks $Q$-SU(2) symmetry and annihilates the tower, the term $h S^z$ acts as $H_\mathrm{SGA}$ and gives energy to the states in the tower, while the term $\propto D$ commutes with $Q^z$ and $Q^{+}$. 
The third neighbour term is added to break a non-local SU(2) symmetry for which the states belonging to the tower are the only states in their symmetry sector.  

\end{shaded}

\subsection{Hilbert space fragmentation}\label{sec:krylov}

Spectrum generating algebra relies on the tower operator $Q^\dagger$ to construct the non-thermalising subspace. We now describe a related mechanism which produces exactly embedded subspaces, but does not require \emph{a priori} knowledge of $Q^\dagger$.  For a Hamiltonian $H$ and some arbitrary vector in the Hilbert space, $|\psi_0\rangle$, the \emph{Krylov subspace}, $\mathcal{S}$, is defined as the set of all vectors obtained by repeated action of $H$ on $|\psi_0\rangle$, 
\begin{equation}
	\mathcal{S}\left(H, \ket{\psi_0}\right) \equiv \textrm{span}\{\ket{\psi_0}, H\ket{\psi_0}, H^2\ket{\psi_0}, \cdots\}.
	\label{eq:krylov}
\end{equation}
Readers familiar with numerical linear algebra will  recall the  same subspace $\mathcal{S}$ is used in iterative methods for finding extremal eigenvalues of large matrices, such as the Arnoldi and Lanczos algorithms.  By definition, $\mathcal{S}$ is closed under the action of $H$.
While $\ket{\psi_0}$ in Eq.~(\ref{eq:krylov}) can in principle be an arbitrary state, in physics  one is primarily interested in initial product states (also called the ``root" states), which are more easily preparable in experiment. 
For a generic non-integrable Hamiltonian $H$ without any symmetries, one expects that $\mathcal{S}\left(H, \ket{\psi_0}\right)$ for \textit{any} initial product state $\ket{\psi_0}$ is the \textit{full} Hilbert space of the system.  If $H$ has some symmetry (and assuming $\ket{\psi_0}$ an eigenstate of the symmetry), $\mathcal{S}\left(H, \ket{\psi_0}\right)$ is expected to span all states with the same symmetry quantum number as $\ket{\psi_0}$. 

Surprisingly, it has been shown that in many cases the system can exhibit   \textit{fragmentation} i.e., even after resolving the symmetries,  $\mathcal{S}\left(H, \ket{\psi_0}\right)$ does \textit{not} span all states with the same symmetry quantum numbers as $\ket{\psi_0}$~\cite{Pai2019,Khemani2019_2, MoudgalyaKrylov, Sala2019}. Following the notations in Ref.~\onlinecite{MoudgalyaKrylov},  we can formally state  this as 
\begin{equation}
\mathcal{H} = \bigoplus_{\bf{s}} \mathcal{H}^{(\bf{s})}, \quad \mathcal{H}^{(\bf{s})} = \bigoplus_{i = 1}^{\#^{(\bf{s})}} {\mathcal{S}\left(H, \ket{\psi_i^{(\bf{s})}}\right)} \, ,
\label{eq:fullhilbert}
\end{equation}
where $\bf{s}$ labels the distinct symmetry quantum numbers,  $\#^{(\bf{s})}$ denotes the number of disjoint Krylov subspaces generated from product states with the same symmetry quantum numbers, and $\ket{\psi_i^{(\bf{s})}}$ are the root states generating the Krylov subspaces. Note that the root states in Eq.~\eqref{eq:fullhilbert} are chosen such that they generate distinct disconnected Krylov subspaces (the same subspace can be generated by different root states). If one (or more) Krylov sectors are non-thermalising, we recognise Eq.~(\ref{eq:fullhilbert}) is of the same form as our previous Eq.~(\ref{Eq:decompose}).

Fragmentation as in Eq.~(\ref{eq:fullhilbert}), where the total number of Krylov subspaces is exponentially large in the system size, was recently shown to \textit{always} exist in Hamiltonians and random-circuit-models with conservation of dipole moment (see Box~4). Before illustrating this for a particular model, we note that, generally, one can distinguish between ``strong" and ``weak" fragmentation, depending respectively on whether or not the ratio of the largest Krylov subspace to the Hilbert space within a given global symmetry sector vanishes in the thermodynamic limit. Strong (resp. weak) fragmentation is associated with the violation of weak (resp. strong) ETH with respect to the full Hilbert space. 
Moreover, different fragments can exhibit vastly different dynamical properties. 
For example, some fragments (even though exponentially large in system size) may be integrable, while others may be non-integrable. Among the non-integrable ones, more subtle ETH-breaking phenomena are also possible~\cite{Moudgalya2019}. 

To illustrate fragmentation, consider the following model of fermions hopping on an open 1D chain while preserving their centre of mass~\cite{MoudgalyaKrylov}: 
\begin{equation}
\label{eq:pairhopping}
H =  \sum_{j} {\left(c_j^\dagger c_{j+3}^\dagger c_{j + 2} c_{j + 1} + \mathrm{h.c.}\right)},
\end{equation}
where $c_j^\dagger$, $c_j$ are the standard fermion creation/annihilation operators on site $j$.  This model has a variety of physical realisations, which we discuss at the end of this section. We now demonstrate the dynamical fragmentation in this model, closely following the presentation in  Ref.~\onlinecite{MoudgalyaKrylov}.

\begin{figure}[t]
	\includegraphics*[width=0.96\linewidth]{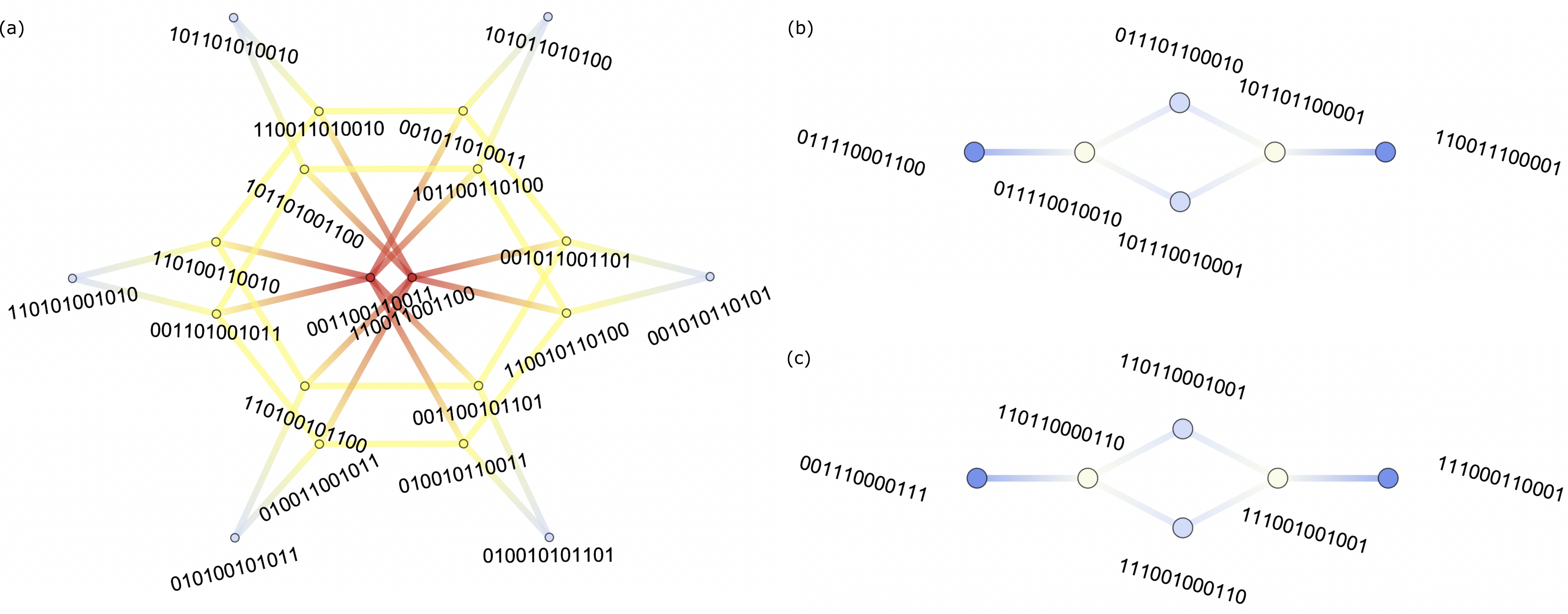}
	\caption{ Hilbert space fragmentation in a pair-hopping  model in Eq.~(\ref{eq:pairhopping}) studied in Ref.~\onlinecite{MoudgalyaKrylov}. Plots (a), (b) and (c) show graphs of three disconnected sectors of the Hilbert space for a small chain with $N{=}6$ electrons at filling factor $\nu{=}1/2$. Each vertex represents a Fock basis state, and vertices are connected by an edge if the Hamiltonian matrix element is non-zero between those basis states. This type of graph is known as the adjacency graph of the Hamiltonian. In this case, because all the matrix elements of the Hamiltonian are equal in magnitude, the adjacency graph is unweighted. The graph has been coloured according to the vertex degree (with red colour indicating high connectivity). 	
	}\label{fig:shatter}
\end{figure}
At filling factor $\nu=1/2$, i.e., with half as many fermions as sites in the chain $L$, it is convenient to split the chain into 2-site cells assuming $L$ is even. For each cell, define the new degrees of freedom:
\begin{eqnarray}
	\ket{\uparrow} \equiv \ket{0 1}\, , \;\;\; \ket{\downarrow} \equiv \ket{1 0} \;\;\; \ket{+} \equiv \ket{1  1}\, , \;\;\; \ket{-} \equiv \ket{0 0}.
	\label{eq:halffillingpart}
\end{eqnarray}
It is fruitful to name these composite degrees of freedom: $\ket{+}$, $\ket{-}$ are called ``fractons", $\ket{+-}$, $\ket{-+}$ are ``dipoles", and $\ket{\uparrow}$, $\ket{\downarrow}$ are ``spins". By inspecting the possible action of the Hamiltonian in Eq.~(\ref{eq:pairhopping}) on these states, we find several types of allowed processes. For example, the following processes can be interpreted as free propagation of dipoles when separated by spins:
\begin{eqnarray}
	\ket{\downarrow + -} &\leftrightarrow& \ket{+ - \downarrow} \, , \label{eq:dipolescattering1} \\
	\ket{- + \uparrow} &\leftrightarrow& \ket{\uparrow - +} \, . \label{eq:dipolescattering2}
\end{eqnarray}
Similarly, the following processes imply that a fracton can only move through the emission or absorption of a dipole:
\begin{eqnarray}
	\ket{\downarrow + \uparrow} &\leftrightarrow& \ket{+ - +} \, , \label{eq:+fracton} \\
	\ket{\uparrow - \downarrow} &\leftrightarrow& \ket{- + -}, \label{eq:-fracton}
\end{eqnarray}
Here, Eqs.~(\ref{eq:dipolescattering1})-(\ref{eq:-fracton}) resemble the rules restricting the mobility of fracton phases of matter~\cite{FractonReview}. However, in contrast to usual fracton phenomenology, here the movement of fractons is also sensitive to the background spin configuration. For example, the fracton in the configuration $\ket{\cdots \downarrow + \uparrow \cdots}$ can move by emitting a dipole (see Eq.~\eqref{eq:+fracton}) while that in the configuration $\ket{\cdots \uparrow + \downarrow \cdots}$ cannot.

The model in Eq.~(\ref{eq:pairhopping}) has several symmetries, most notably the total charge (number of fermions) and the total dipole moment are conserved,
\begin{equation}
	\widehat{Q}=\sum_j \widehat{Q}_j, \quad \quad 	\widehat{D} \equiv \sum_{j} j \widehat{Q}_j,
	\label{eq:dipoleoperator}
\end{equation}
with $\widehat{Q}_j \equiv \hat{n}_{2j - 1} + \hat{n}_{2j} - 1$,  where $j$ is the unit cell index and $2j-1$, $2j$ are the site indices of the original configuration. 

Exponentially many of Krylov subspaces in the model in Eq.~(\ref{eq:pairhopping}) are one-dimensional frozen  configurations. 
For instance, the Hamiltonian vanishes on any product state that does not contain the patterns $``\cdots 0110 \cdots"$ or $``\cdots 1001\cdots"$, since those are the only configurations on which terms of $H$ act non-trivially. The state 
\begin{equation*}
	\ket{1111000011110000\dots\dots1111000011110000}
\end{equation*}
is one example of a static configuration that is an eigenstate.
In terms of the composite degrees of freedom we can equivalently consider configurations with only $+$, $-$, and no spins, such as $$\ket{\cdots + + - - + + - - \cdots },$$ with a pattern that alternates between $+$ and $-$ with `domain walls' that are at least 2  sites apart. Once again, it is easy to see that all terms of the Hamiltonian vanish on these configurations: since there are \textit{exponentially} many such patterns, there are equally many one-dimensional Krylov subspaces.

In Ref.~\onlinecite{MoudgalyaKrylov} it was found that large fragments (with dimension exponential in system size $L$) can be integrable (mappable to spin-$1/2$ XX spin model) as well as non-integrable. See Fig.~\ref{fig:shatter} for an illustration of some of the disconnected sectors of the Hilbert space. The dynamics initialised in any of the states, e.g., in Fig.~\ref{fig:shatter}(a) cannot reach any of the states in plots (b) and (c).

We previously announced that the pair-hopping model in Eq.~(\ref{eq:pairhopping}) arises in a number of different physical contexts:  (i) it arises as the dominant hopping process for electrons in the regime of the fractional quantum Hall effect in a quasi-1D limit~\cite{Moudgalya2019}; (ii) it arises in the Wannier-Stark problem, i.e., spinless fermions hopping on a one-dimensional lattice, subject to a large electric field~\cite{wannierrmp}; (iii) it can be mapped to the following spin-1 fractonic model studied in Ref.~\cite{Sala2019}: $ H_4= -\sum_{n}\left( S_n^+S_{n+1}^-S_{n+2}^-S_{n+3}^++\text{h.c.}\right)$; (iv) at filling factor $\nu{=}1/3$, one of the Krylov sectors can be mapped to the so-called PXP model, introduced in Eq.~(\ref{Eq:PXP}) below, which describes a chain of strongly-interacting Rydberg atoms. Note that in realisations (i), (ii) there are usually additional diagonal terms in the Hamiltonian that are of the same order as the hopping. While this does not affect the fragmentation, it may significantly impact other dynamical properties within the fragments. The realisation (ii) has recently been investigated as a platform for many-body localisation without disorder~\cite{StarkMBL1,StarkMBL2}. 

Signatures of fragmentation have been observed in other models including, e.g., the Fermi-Hubbard model and its cousins~\cite{Vafek,  IadecolaZnidaric}, various constrained models~\cite{Sikora2011, Lan2017_2, Gopalakrishnan2018}, and bosons in optical lattices~\cite{bosonScars, Zhao2020}. In the latter case, the Krylov subspaces are only \emph{approximately} exact in the sense that there exist non-zero matrix elements which connect different Krylov subspaces but their magnitude is much smaller than the matrix elements within a given Krylov subspace.

\definecolor{shadecolor}{rgb}{0.8,0.8,0.8}
\begin{shaded}
	\noindent{\bf Box 4 $|$ Fragmentation in quantum circuits}
\end{shaded}
\vspace{-9mm}
\definecolor{shadecolor}{rgb}{0.9,0.9,0.9}
\begin{shaded}

\noindent Beyond Hamiltonian systems, fragmentation generally arises in models of  1D random unitary circuit dynamics, constrained to conserve both U(1) charge and its dipole moment~\cite{Pai2019,Khemani2019_2}.  Consider a model from Ref.~\onlinecite{Pai2019} with a chain of $S{=}1$ quantum spins, with the  local $z$-basis $|+\rangle, |-\rangle, |0\rangle$, and unitary gates which locally conserve charge $\widehat{Q} = \sum_j S_j^z$ and dipole moment $\widehat{D} = \sum_{ j} j S^z_j$.  These intertwined conservation laws greatly restrict the allowed movement of charges, e.g., a single $+$ or $-$ charge on site $x$ has dipole moment $D{=} \pm x$. Such a charge cannot hop to the left or right, because this would change the net dipole moment by one unit. On the other hand, bound states of charges  or ``dipoles" of the form $(-+)$ have net charge zero and net dipole moment $D{=}\pm1$ independent of position, and these can move freely. Additionally, dipoles can enable the movement of charges, because a charge can move if it simultaneously emits a dipole to keep $D$ unchanged: $|0+0\rangle \rightarrow |+-+\rangle$. 

The above phenomenology is very similar to the pair hopping model discussed in the main text, but following Ref.~\cite{Khemani2019_2},  we realise it  using Floquet circuits composed from $\ell$-site unitary gates, which locally conserve $Q$ and $D$. These $\ell$-site gates are therefore block-diagonal, and they are applied periodically in time.  Assuming for simplicity $\ell{=}3$, we tile the chain by 3-site unitaries, staggered across three layers, i.e., $ U^F = U^1 U^2 U^3$, with $ U^1 \equiv U^1_{1,2,3}  U^1_{4,5,6} \cdots$ illustrated in blue colour in Fig.~B4 below (and similarly for $U^2$, $U^3$ in red and green colour, respectively).  The complete circuit consists of repeatedly applying $U^F$ some number of times. Note that  $U^{1}, U^{2}$ and $U^{3}$ can be chosen at random for a given realisation, but then remain fixed throughout the circuit. 

\begin{minipage}[c]{\linewidth}
	\begin{center}
		\includegraphics[width=0.4\linewidth]{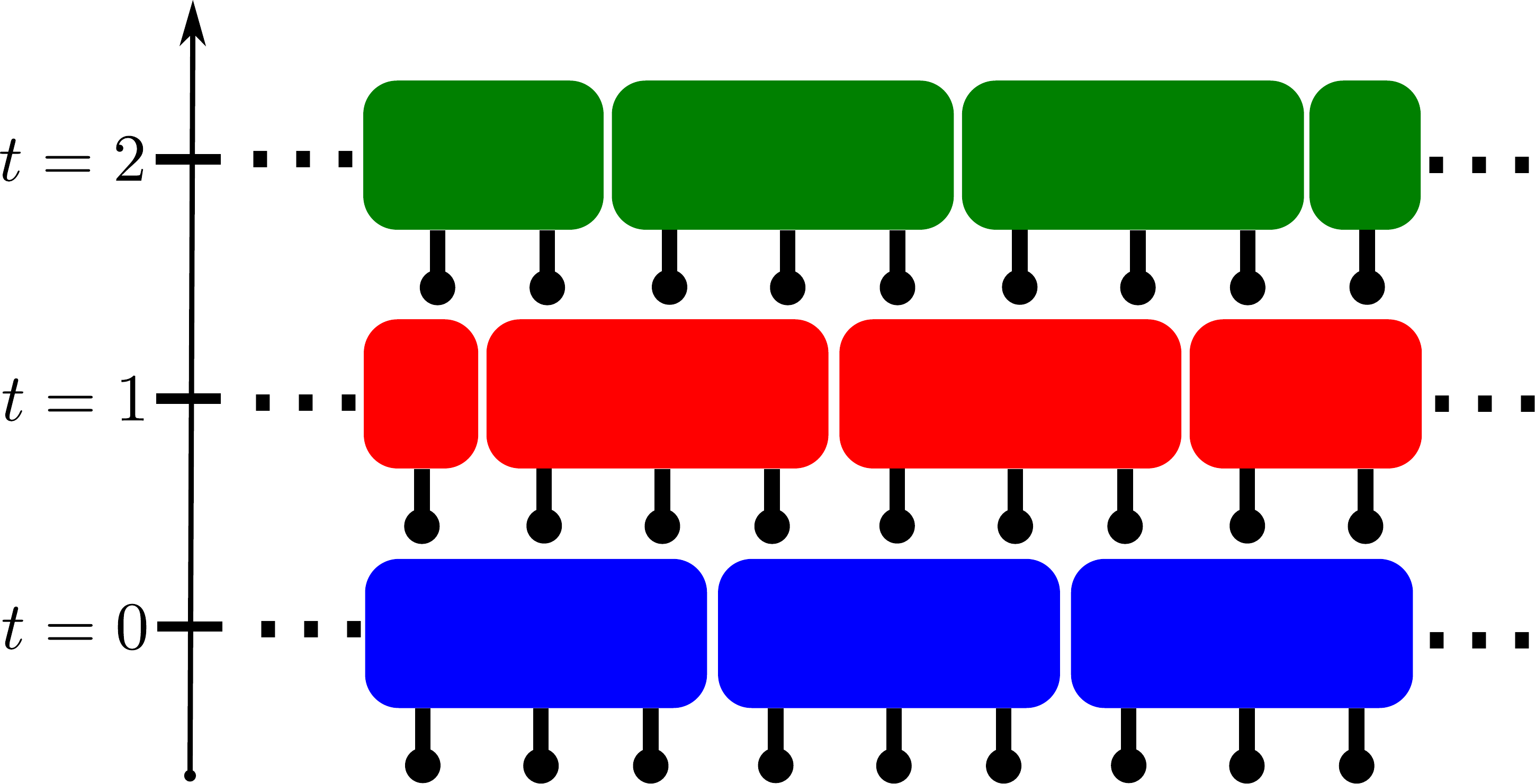}	
	\end{center}
	\begin{center}
		{\small {\bf Figure B4:} Illustration of the Floquet operator $ U^F = U^1 U^2 U^3$, 
			staggered across three layers from Ref.~\onlinecite{Khemani2019_2}. }
	\end{center}
	
\end{minipage}
\vspace*{0.2cm}

Ref.~\onlinecite{Khemani2019_2} showed that local fractonic circuits of this type must have exponentially many Krylov sectors. To see this, note that {\it any} pattern that alternates between locally `all plus' and locally `all minus,' with domain walls in between at least $\ell$ sites apart, must be inert. These are states of the form 
$	|++++-----++++ \cdots\rangle$. The argument is similar to the one given in the main text for the Hamiltonian system. 
One can then  lower bound the size of the disconnected subspace by dividing the system up into blocks of length $\ell$, and allowing each block to be either `all plus' or `all minus.' This yields an inert subspace of dimension at least $2^{L/\ell} = c^L$, where $c = 2^{1/\ell}$. This bound is not tight~\cite{Khemani2019_2}, but it proves the existence of an exponentially large, localised subspace for any finite gate size.  We emphasise that the exponentially large number of sectors goes beyond simple symmetry considerations, which would only predict $\propto L^3$ sectors, since the allowed values of quantum numbers are $-L \leq  Q \leq  L$ and $-L(L-1)/2 \leq D \leq  L(L-1)/2$, where $L$ is the length of the chain.

\end{shaded}

\subsection{Projector embedding}

At the highest level of abstraction, we can ask if one could embed an arbitrary subspace into the spectrum of a thermalising system? For concreteness, assume we are given an arbitrary set of states $ \vert \psi_i \rangle$ that span our target non-thermalising subspace $H_{\mathrm{non-ETH}}=\mathrm{span}\{  \vert \psi_i \rangle\}$, which we wish to embed into a thermalising Hamiltonian $H$ as in Eq.~(\ref{Eq:decompose}). 
This can be achieved via the ``projector embedding" construction first introduced by Shiraishi and Mori~\cite{ShiraishiMori}. 

Our target states $\vert \psi_i \rangle$ are assumed to be non-thermal, hence we can furthermore assume there exists a set of  local projectors, $P_i$, which annihilate these states, $ P_i \vert \psi_j \rangle = 0$, for any $i$ ranging over lattice sites $1,2,\ldots,L$.  Next, consider a lattice Hamiltonian of the form
\begin{eqnarray}\label{eq:shiraishi}
H = \sum_{i=1}^L P_i h_i P_i + H^{\prime},
\end{eqnarray}
where  $h_i$ are arbitrary operators that have support on a finite number of sites around $i$, and $[H^{\prime},P_i] = 0$ for all  $i$. It follows
\begin{eqnarray}
P_i H \vert \psi_j \rangle = P_i H^{\prime} \vert \psi_j \rangle = H^{\prime} P_i \vert \psi_j \rangle = 0,
\end{eqnarray}
thus $[H,\sum_i P_i]=0$. Therefore, $H$ takes the desired block diagonal form in Eq.~(\ref{Eq:decompose}).
The target eigenstates can in principle be embedded at arbitrarily high energies for a suitably chosen $H'$, which also ensures that the model is overall non-integrable~\cite{ShiraishiMori}. However, there is no guarantee that the embedded states must be equidistant in energy and they may even be degenerate, such that this scheme could clearly result in a model without an SGA property discussed in Sec.~\ref{sec:sga}.

Physical applications of projector embedding include various topologically ordered systems~\cite{NeupertScars,Wildeboer2020} and models with lattice supersymmetry~\cite{Surace2020}, which are themselves defined in terms of local projectors (see Box~5). 
While the projector construction allows to embed general classes of states into a thermalising spectrum, in practice one is often more interested in the reverse question: given a Hamiltonian $H$ belonging to some known physical system, can one identify non-thermal embedded states $ \vert \psi_i \rangle$? Although this task is obviously much harder, for some physical models such as the PXP model, introduced in Sec.~\ref{sec:pxp} below, this question has been answered affirmatively for a single target state at zero energy in the middle of the spectrum, which has been identified with the AKLT ground state in Eq.~(\ref{eq:AKLTMPS})~\cite{Shiraishi_2019}. However, for the AKLT non-thermal eigenstates constructed via the SGA in Sec.~\ref{sec:sga}, in particular for the tower of states in Eqs.~(\ref{eq:tower})-(\ref{eq:aklttower}), at present it is not known whether it is possible to find a set of local projectors and cast the entire tower of AKLT states in the Shiraishi-Mori form. This suggests there may be broader classes of models with non-thermal eigenstates that extend beyond the form in Eq.~(\ref{eq:shiraishi}). Indeed, in a very recent work~\cite{Langlett2021} a scheme based on a double copy of the system has been proposed for embedding eigenstates with high entanglement.

\definecolor{shadecolor}{rgb}{0.8,0.8,0.8}
\begin{shaded}
	\noindent{\bf Box 5 $|$ Physical realisations of projector embedding}
\end{shaded}
\vspace{-9mm}
\definecolor{shadecolor}{rgb}{0.9,0.9,0.9}
\begin{shaded}
	
\noindent To illustrate constructions of non-thermal eigenstates via projector embedding, we  consider a class of topological lattice models defined in terms of local projectors, following Ref.~\onlinecite{NeupertScars}. A typical Hamiltonian of such models has the form
	\begin{align}
		H(\beta)\:=
		\sum_{p}
		\alpha^{\,}_{p}\, Q^{\,}_{p}(\beta),
		\label{eq:1a Hamiltonian}
	\end{align}
	where $p$ labels, e.g., elementary plaquettes of a lattice, such as in toric code~\cite{KitaevToricCode}. The operators
	$Q^{\,}_{p}(\beta)$ are Hermitian, positive-semidefinite, and local. They contain only sums of products of operators defined within the bounded region
	labeled by $p$, e.g., at the Rokhsar-Kivelson point of the quantum dimer model on the square lattice~\cite{RokhsarKivelson},  $Q^{\,}_{p}(\beta)$ are	projectors that encode both the potential and kinetic (plaquette flip) terms. 
	The dimensionless parameter $\beta$ is used to deform 
	solvable models and break integrability. 
    
    The operators $Q^{\,}_{p}(\beta)$
	are built so as to share a common null state
	$|\Psi(\beta)\rangle$, i.e., for every $p$ we have   $Q^{\,}_{p}(\beta)\;|\Psi(\beta)\rangle=0$.  If all the couplings $\alpha^{\,}_{p}$ are
positive, the state $|\Psi(\beta)\rangle$ is the ground state of $H(\beta)$. If,  instead,  $ \alpha^{\,}_{p}$ take both positive and negative values, then it is not guaranteed that $|\Psi(\beta)\rangle$ is a ground state. Nevertheless, $|\Psi(\beta)\rangle$ is still an eigenstate with energy $E{=}0$. When this state is a high-energy eigenstate  of $H(\beta)$, it is an
atypical state in that it displays area law entanglement entropy, since it is also a ground state of a \emph{different} local Hamiltonian:
\begin{eqnarray}
H^\prime (\beta)\:=\sum_{p} |\alpha^{\,}_{p}|\, Q^{\,}_{p}(\beta).
\end{eqnarray}
Hence $|\Psi(\beta)\rangle$ is a non-thermal state, if $H(\beta)$ is nonintegrable.  By deforming exactly solvable models -- the toric code, for instance -- one can break integrability while retaining the $E{=}0$ state~\cite{NeupertScars}. The construction outlined here can be shown to be equivalent to the embedding construction in Eq.~(\ref{eq:shiraishi}) by diagonalising the operators $Q^{\,}_{p}(\beta)$ and separating the terms belonging to zero eigenvalues (which must exist by construction) from other eigenvalues.

\end{shaded}

\section{PXP model}\label{sec:pxp}

In the previous section, we introduced several mechanisms of weak ergodicity breaking, realised by diverse physical systems. While these mechanisms are mutually independent and different systems may only exhibit some of them, we next show how these mechanisms ``co-exist" within a paradigmatic model of weak ergodicity breaking known as the PXP model. This model arises as an effective description of Rydberg atoms in the regime of a strong Rydberg blockade, and its experimental realisation will be the topic of  Sec.~\ref{sec:exp}. As we explain below, the PXP model realises much of the weak ergodicity breaking phenomenology discussed previously, in particular its Hilbert space is fragmented and it contains non-thermal eigenstates which form an approximate SGA.  

\subsection{The model}

The  PXP model describes a chain of coupled two-level systems, where each system can be in one of the two possible states, $\ket{\circ}$  and $\ket{\bullet}$. In the Rydberg atom realisation, these states correspond to an atom being in its ground state or in the excited Rydberg state, respectively. However, for present purposes, we can view these two states as $\downarrow$, $\uparrow$ projections of a spin-1/2 degree on the given site.  We consider a 1D chain of such two-level systems coupled according to  the ``PXP" Hamiltonian~\cite{Lesanovsky2012}, 
\begin{equation}\label{Eq:PXP}
	H_{\mathrm{PXP}} = \sum_i P_{i-1}\sigma^x_i P_{i+1},
\end{equation}
where $\sigma_i^x \equiv \ket{\circ_i}\bra{\bullet_i} + \ket{\bullet_i}\bra{\circ_i} $ is the standard Pauli $x$-matrix on site $i$ and $P_i  = \ket{\circ_i}\bra{\circ_i}$ is the projector on the ground state at site $i$. Equivalently, the projectors can be defined as $P_i = \frac{1}{2}(1-\sigma_i^z)$, with $\sigma_i^z \equiv \ket{\circ_i}\bra{\circ_i} - \ket{\bullet_i}\bra{\bullet_i} $ . 

Without projectors, the Hamiltonian in Eq.~(\ref{Eq:PXP}) is that of a free paramagnet: each spin would independently precess with the Rabi frequency set to 1. $P$  introduces a kinetic constraint: it allows a spin to flip only if both of its nearest neighbours are in $\circ$ state. For example, the process $\cdots \circ\circ\circ\cdots \leftrightarrow \cdots \circ\bullet\circ\cdots$ is allowed, while $\cdots \bullet\circ\circ\cdots \leftrightarrow \cdots \bullet\bullet\circ\cdots$ is forbidden. This makes the model intrinsically interacting, as it is no longer possible to describe the state of each spin independently of other spins. Numerical simulations based on exact diagonalisation of the PXP model with up to $L{=}32$ spins have demonstrated that the statistics of its energy-level spacings approached the prediction of random matrix theory~\cite{Turner2017} as the system size was increased. Hence, despite its very simple form, we expect the PXP model cannot be fully ``solved" using the known integrability techniques~\cite{Sutherland}.

Another consequence of projectors is the fragmentation of the PXP Hamiltonian: $H_\mathrm{PXP}$ splits into sectors corresponding to different numbers of adjacent spin excitations. The largest connected component of the Hilbert space is one that excludes any configurations with adjacent excitations, $\ldots {\bullet}{\bullet} \ldots$. The number of classical configurations that satisfy such a constraint is still exponentially large -- more precisely, it scales asymptotically 
\begin{eqnarray}
{\cal D}_L\propto \varphi^L, \quad \varphi = \frac{1+\sqrt{5}}{2},
\end{eqnarray}
where $\varphi$ is  the golden ratio. This unusual scaling is a manifestation of the global constraint. Similar type of constraints can arise due to emergent gauge fields and have been used to model interactions between anyon excitations in topological phases of matter~\cite{Feiguin07}. Apart from this largest sector, there are further sectors which contain some number of nearest neighbour spin flips; however, in the remainder of this section we will focus on the \emph{largest} connected component of the Hilbert space, which already displays non-trivial weak ergodicity breaking phenomenology.

\subsection{Ergodicity breaking in the PXP model}

\begin{figure}[t]
	\includegraphics[width=0.98\linewidth]{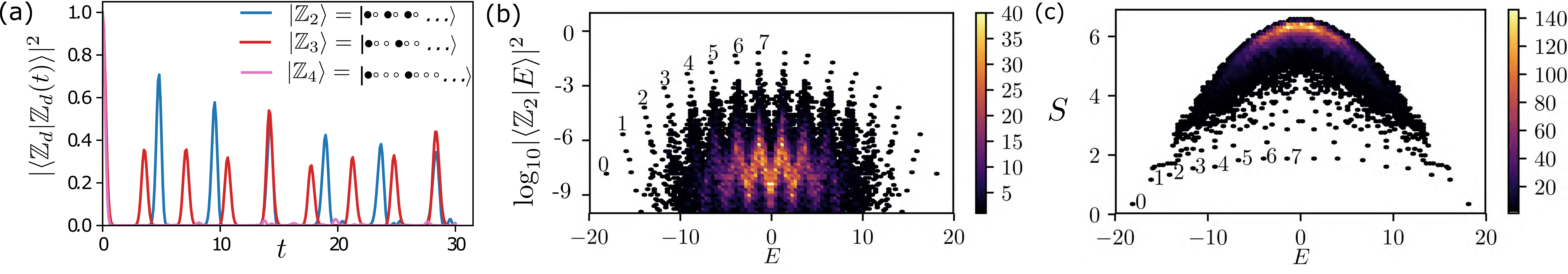}
	\caption{
		(a) Numerical simulation of quantum fidelity in a large PXP model with $L{=}30$ atoms reveals strong atypicality with respect to the initial state~\cite{TurnerPRB}. For period-2 and -3 density waves, the fidelity features robust revivals, while period-4 density wave shows no revivals. 
		(b)-(c): Non-thermal eigenstates violate the ETH by their anomalously large projection on $|\mathbb{Z}_2\rangle$ state, and by their low entanglement entropy, $S$~\cite{TurnerPRB}.  Each point represents a single eigenstate with energy $E$ in a large PXP chain with $L{=}30$ atoms. Scarred eigenstates have been numbered by $0,1,\ldots,7$. Colour scale represents the density of data points.	
		Panels reproduced with permission from Ref.~\onlinecite{TurnerPRB} APS, under a Creative Commons licence CC BY 4.0.
	} \label{fig:pxp}
\end{figure}

While the PXP model is quantum-chaotic, numerical simulations of its quench dynamics ~\cite{Sun2008, LesanovskyDynamics, Turner2017, TurnerPRB} have revealed surprising non-ergodic behaviour. For example,  the return probability in global quenches with the PXP Hamiltonian is shown in Fig.~\ref{fig:pxp}(a)~\cite{Turner2017,TurnerPRB}. `Global quench' means that the system is prepared in a highly non-equilibrium initial state, $\ket{\psi_0}$, at time $t{=}0$, and the system is subsequently evolved with the many-body Hamiltonian, $H_\mathrm{PXP}$ in Eq.~(\ref{Eq:PXP}). Since the PXP Hamiltonian is purely off-diagonal in the standard $z$-basis, any classical product state has an average energy equal to zero and extensive energy variance, thus effectively playing the role of an ``infinite temperature" ensemble. 
In Fig.~\ref{fig:pxp}(a), three choices of density-wave states were considered for $\ket{\psi_0}$: $\ket{\mathbb{Z}_2}\equiv \ket{\cdots {\bullet}{\circ}{\bullet}{\circ}\cdots}$, $\ket{\mathbb{Z}_3}\equiv \ket{\cdots {\bullet}{\circ}{\circ}{\bullet}{\circ}{\circ}\cdots}$ and $\ket{\mathbb{Z}_4}\equiv \ket{\cdots {\bullet}{\circ}{\circ}{\circ}{\bullet}{\circ}{\circ}{\circ}\cdots}$. We note that these states can be prepared in Rydberg atom experiments by modulating the so-called detuning term introduced below in Eq.~(\ref{eq:rydberg}).

As the system evolves following the quench, one can characterise its behaviour by the return probability, also known as many-body fidelity, which quantifies the probability of observing the initial state after unitary dynamics, 
\begin{equation}\label{Eq:fidelity}
	F(t) = |\braket{\psi_0 | \psi(t)}|^2 =  |\bra{\psi_0} e^{-iHt} \ket{\psi_0}|^2.
\end{equation}
Intuitively, thermalising dynamics leads to a quick spreading of the many-body wave function over the full Hilbert space. According to  Fig.~\ref{Fig1}(b), fidelity is expected to rapidly decrease to an exponentially small value, $F\propto 1/{\cal D}_L$, and remain near that value at late times. The numerical result in Fig.~\ref{fig:pxp}(a) defies this expectation for \emph{certain} initial states. In particular, the $\mathbb{Z}_2$ and $\mathbb{Z}_3$  density wave states show pronounced revivals at certain times, $F(nT) \sim O(1)$. In contrast, quench dynamics from $\mathbb{Z}_4$  (and many other initial product states not shown) displays fast relaxation without revivals.  We note that the dynamics of local observables, such as the density of domain walls nucleated in $\mathbb{Z}_2$, i.e., the number of $\cdots {\circ}{\circ}\cdots$ patterns, is very similar to the fidelity dynamics. In particular, the frequency of revivals in local observables is the same as that of fidelity for the initial state $|\mathbb{Z}_2\rangle$. Below we focus on understanding the dynamics for this initial state.

In order to understand the origin of the atypical dynamical behaviour in the PXP model, we write the return amplitude as  
\begin{eqnarray}
\langle \mathbb{Z}_2|e^{-i t H}|\mathbb{Z}_2\rangle = \sum_n e^{-i t E_n}|\langle E_n |\mathbb{Z}_2\rangle|^2,
\end{eqnarray}
where $E_n$, $|E_n\rangle$ denote the eigenenergies and eigenvectors of $H_{\mathrm{PXP}}$, respectively. Thus, quantum evolution is fully determined by the eigenenergies, $\{ E_n \}$, and overlaps of energy eigenstates with the initial state, $\{ |\langle E_n |\mathbb{Z}_2\rangle|^2\}$. Any special features in the dynamics translate into atypical properties of these overlaps. Indeed, Fig.~\ref{fig:pxp}(b) reveals a set of eigenstates that have strongly enhanced overlaps with $\mathbb{Z}_2$ state, $|\langle E |\mathbb{Z}_2\rangle|^2$. These eigenstates violate the ETH~\cite{DeutschETH,SrednickiETH,RigolNature}, since we expect that individual eigenstates of thermalising systems should behave like thermal ensembles.  By contrast, Fig.~\ref{fig:pxp}(b) demonstrates that special eigenstates have strongly enhanced overlaps with a particular product state, thus they are anomalously concentrated in the Hilbert space and do not resemble random vectors. 

Note that violations of the ETH are traditionally probed by studying the matrix elements of local observables~\cite{Rigol07}. The overlaps with $|\mathbb{Z}_2\rangle$ state indeed probe the matrix elements of the alternating magnetic field operator, $\sum_j (-1)^j \sigma_j^z$, for which  $|\mathbb{Z}_2\rangle$ is the highest weight state.  The higher the overlap with $|\mathbb{Z}_2\rangle$, the stronger the violation of the ETH -- thus, the special eigenstates are the ``most"  ETH violating states in the spectrum of the PXP model. Intriguingly, the highest overlaps with $|\mathbb{Z}_2\rangle$ are achieved in the \emph{middle} of the spectrum, i.e., at the highest effective temperature in this system. 

The total number of special states in Fig.~\ref{fig:pxp}(b) scales with the number of spins as $L{+}1$~\cite{TurnerPRB}. Moreover,  the special states are equidistant in energy, especially near the middle of the spectrum. Along with their large overlap with $|\mathbb{Z}_2\rangle$ state, this explains the existence of fidelity revivals. However, as seen from the coloured density of data points in Fig.~\ref{fig:pxp}(b), while the majority of other states in the spectrum have vanishingly small overlap with $\ket{\mathbb{Z}_2}$, there is still a considerable number of states forming ``towers" which cluster around the energy of the special eigenstates. Towers of scarred eigenstates are also found in the PXP model in the presence of periodically driven detuning~\cite{Mukherjee2020}, and in a two-dimensional PXP model~\cite{Michailidis2020}, where the overlap is computed with a charge-density wave state having a checkerboard pattern~\cite{Michailidis2D,Hsieh2020}. 

Furthermore, the same $L{+}1$ special eigenstates show anomalously low entanglement entropy as seen in Fig.~\ref{fig:pxp}(c). For a thermalising eigenstate, the bipartite entanglement entropy of a subsystem of length $L_A$ is expected to scale extensively with the volume of the subsystem, $S\propto L_A$. In contrast, the entanglement entropy of special eigenstates is found to scale approximately with the logarithm of the subsystem size, as in Eq.~(\ref{eq:logscaling})~\cite{TurnerPRB}. This is another evidence of ETH violation as it shows that special eigenstates are not uniformly spread over the Hilbert space. 

Recall that logarithmic scaling of entanglement entropy is commonly found in other models with SGA, Eq.~(\ref{eq:logscaling}). Indeed, approximations to special eigenstates can be constructed starting from an approximate ground state of the PXP model and creating spin wave excitations on top of it~\cite{Iadecola2019}. Although the accuracy of this scheme deteriorates for special eigenstates near the middle of the spectrum, the scheme provides an intuitive picture of special eigenstates as condensates of weakly-interacting magnons, similar to models with an SGA.  An alternative approach in Ref.~\cite{lin2018exact} focused on the middle of the energy spectrum of the PXP model, where a few exact eigenstates can be constructed, thus rigorously demonstrating the ETH violation (see Box~6).  While exact MPS constructions similar to Ref.~\onlinecite{lin2018exact} can be generalised to other models, e.g., a  transverse Ising ladder~\cite{Voorden2020} and the Floquet versions of the PXP model~\cite{Sugiura2019,Mizuta2020}, they remain limited to a small number of states in the middle of the spectrum, and in particular they do not capture the top states in the towers, which are not simple MPS due to having larger-than-area-law entanglement.

Finally, we note that the model in Eq.~(\ref{Eq:PXP}) with the addition of a few diagonal terms can be made integrable~\cite{FendleySachdev, Feiguin07, LesanovskyMPS} and it has a very rich phase diagram in its the ground state. However, these terms that render the PXP model integrable have magnitude $O(1)$, i.e., they are strong perturbations and are not expected to be relevant for explaining the physics presented in this section. Furthermore, several studies explored the effect of (generally, weaker)  perturbations on the PXP model~~\cite{TurnerPRB, Lin2020, SuracePert}. While non-thermal eigenstates are generally not robust to perturbations, some aspects of atypical dynamical behaviour are found to persist, including in the presence of disorder~\cite{MondragonShem2020}. Finally, we mention that similar phenomenology of weak ergodicity breaking has also been found in the two-dimensional PXP model~\cite{Michailidis2D,Hsieh2020}, and in related models of transverse Ising ladders~\cite{Voorden2020} and the periodically driven PXP model~\cite{Sugiura2019,Mizuta2020, Mukherjee2020}.

\definecolor{shadecolor}{rgb}{0.8,0.8,0.8}
\begin{shaded}
	\noindent{\bf Box 6 $|$ Ergodicity breaking in null spaces}
\end{shaded}
\vspace{-9mm}
\definecolor{shadecolor}{rgb}{0.9,0.9,0.9}
\begin{shaded}
	
	\noindent Among the special properties of the PXP model is the existence of an exponentially large number of states which are exactly annihilated by the PXP Hamiltonian in Eq.~(\ref{Eq:PXP}). This null space contains states with energy $E{=}0$ that reside in the middle of the spectrum~\cite{Turner2017, TurnerPRB, Iadecola2018}. This exponential zero-energy degeneracy is a consequence of the combined action of particle-hole transformation $\mathcal{C} = \prod_i \sigma_i^z$, which \emph{anticommutes} with the PXP Hamiltonian, and spatial inversion which maps site $j$ to $L{-}  j {+} 1$~\cite{Sutherland86,Inui}. 
	
	Among the exponentially many $E{=}0$ eigenstates in the PXP model, there are also a few weakly entangled states which have been analytically constructed by Lin and Motrunich~\cite{lin2018exact}. Such states are compactly written as MPS  in Eq.~(\ref{eq:mpspbc}) with $\chi{=}2$:
	\begin{eqnarray}\label{eq:pxpzeromode}
		A^{\circ\circ} = \begin{pmatrix}
			0 & -1 \\
			1 & 0
		\end{pmatrix}, \;\;
		A^{\circ\bullet} = \begin{pmatrix}
			\sqrt{2} & 0 \\
			0 & 0 
		\end{pmatrix}, \;\;
		A^{\bullet\circ} = \begin{pmatrix}
			0 & 0 \\
			0 & -\sqrt{2}
		\end{pmatrix}. \;\;\;\;\;
		\label{eq:Amatr}
	\end{eqnarray}
	The local Hilbert space on which these matrices are defined is a 2-site block of the chain; due to the constraint, this block can be in three states, $(\circ\circ)$, $(\bullet\circ)$, and $(\bullet\circ)$.  The constraint automatically prevents  configurations with $(\circ\bullet)(\bullet\circ)$ on consecutive blocks, since  
	$A^{\circ\bullet} A^{\bullet\circ} = 0$. 
	Using the block representation, it can be proven~\cite{lin2018exact} that such an MPS state is exactly annihilated by the PXP Hamiltonian in Eq.~(\ref{Eq:PXP}). 
	
	In fact, from the state defined in Eq.~(\ref{eq:pxpzeromode}) we can construct its partner translated by one site, also at $E{=0}$. These two states thus display translational symmetry breaking with a period-2 bond-centered pattern, despite being in 1D and at an infinite temperature. They also manifestly violate the ETH via their entanglement entropy, which must obey the area law due to their explicit MPS form. We note that Ref.~\onlinecite{Surace2021} has recently shown that MPS states such as the one in Eq.~(\ref{eq:pxpzeromode}) can arise more generally in random quantum networks due to connectivity bottlenecks.  
	
	In Ref.~\onlinecite{Karle2021}, a more general exploration of null spaces 
	was performed by applying a ``subspace disentangling" algorithm in order to construct the least entangled state belonging to $E{=}0$ subspace. It was numerically demonstrated on several examples that this least entangled null state obeys area-law entanglement scaling, and as such, violates  strong ETH. 
	Thus, although protected by symmetry, null spaces provide another non-trivial mechanism for realising non-thermal states.  
	
\end{shaded}

\subsection{The origin of non-thermal eigenstates and quantum revivals}

In quantum physics, the simplest system that exhibits non-trivial dynamics and revivals is an elementary spin: a spin pointing along the $z$-direction precesses when the magnetic field is turned on in the $x$-direction, causing  the spin to periodically return to its initial orientation as time passes.   The revivals in the PXP model can also be understood as precession of a spin with magnitude $s{=}L/2$. The latter spin, however, is a collective degree of freedom representing the system of $L$ atoms~\cite{Choi2018}.  More precisely, the $L{+}1$ scarred eigenstates identified in Fig.~\ref{fig:pxp} form an approximate representation of an su(2) algebra for spin-$L/2$.  This perspective brings the PXP model in line with other models discussed in Sec.~\ref{sec:sga}, with the main difference being that the SGA in the PXP model is only \emph{approximate}. 

Specifically, the spin picture follows from the mapping of the PXP model to a tight-binding chain -- see Box~7. For the initial $|\mathbb{Z}_2\rangle$ state, the ``big spin" raising operator 
\begin{eqnarray}\label{eq:hplus}
H^+ = \sum_{i\in {\rm even}} \tilde\sigma^+_i + \sum_{i\in {\rm odd}}\tilde\sigma^-_i,
\end{eqnarray}
excites an atom anywhere on the even sublattice and deexcites an atom on the odd sublattice, where $\tilde\sigma^\pm_i=P_{i-1}\sigma^\pm_iP_{i+1}$ are the on-site raising and lowering operators that respect the constraint~\cite{Turner2017}.  Similarly, the spin lowering operator $H^-$ performs the same process with the sublattices exchanged. 

The reason for this choice of $H^\pm$ is that their commutator defines the $z$-projection of spin,  $H^z  \equiv \frac{1}{2} [H^+, H^-]$, for which $|\mathbb{Z}_2\rangle$ plays the role of the extremal weight state. 
Now the analogy with spin precession is almost complete  because the PXP Hamiltonian in Eq.~(\ref{Eq:PXP}) is given by the sum $H_{\mathrm{PXP}} = H^+ + H^-$, i.e., it plays the role of an $x$-component of spin.  Thus, preparing the atoms in $|\mathbb{Z}_2\rangle$ state is equivalent to initialising the spin along the $z$-axis, and the state revives because the PXP Hamiltonian acts like a transverse magnetic field.

If the above spin picture were exact, the revivals in the PXP model would be perfect, with fidelity in Eq.~(\ref{Eq:fidelity}) reaching 1 at certain late times. This is not seen, either in experiments discussed below or in the numerical simulations of the PXP model -- recall Fig.~\ref{fig:pxp}(a). The reason is that the mentioned su(2) spin algebra is only approximate, $  \left[ H^z, H^{\pm} \right] \approx \pm H^{\pm}$~\cite{Choi2018,Bull2020}, where $``\approx"$ means there are additional operators on the right-hand side with smaller numerical prefactors from the leading $H^\pm$.  

It has been realised that the structure of the su(2) algebra and the robustness of the revivals can be significantly improved by small deformations of the PXP model~\cite{Khemani2018, Choi2018}.  The inclusion of deformations completely arrests the entanglement growth, resulting in the band of scarred eigenstates nearly fully separated from the rest of the spectrum. We note that a similar procedure can be used to enhance revivals from other initial states, such as $\ket{\mathbb{Z}_3}$ and $\ket{ \mathbb{Z}_4}$, for suitably redefined $H^\pm$ operators~\cite{Bull2020}. Thus, the PXP model could be deformed to stabilise different embedded su(2) algebras. The deformations of the PXP model that enhance the su(2) algebra and improve the $\ket{\mathbb{Z}_2}$ revival fidelity suggest the existence of an idealised parent model that hosts ``perfect" many-body scars.  Unfortunately, the parent model is defined by a long-ranged Hamiltonian with  exponentially-decaying tails~\cite{Choi2018}, and it remains unknown whether it can be expressed in a more compact (short-range) form.

\definecolor{shadecolor}{rgb}{0.8,0.8,0.8}
\begin{shaded}
	\noindent{\bf Box 7 $|$ Forward scattering approximation}\label{box:fsa}
\end{shaded}
\vspace{-9mm}
\definecolor{shadecolor}{rgb}{0.9,0.9,0.9}
\begin{shaded}
\noindent A free paramagnet is an example of a perfectly reviving system: any product state of spins pointing along the $z$-axis exhibits periodic dynamics under the magnetic field along the $x$-direction. Alternatively, the free paramagnet can be viewed as a hypercube graph of dimension $L$, where each of the $2^L$ spin product states represents a vertex, while the edges connect vertices that can be reached by flipping one spin. This picture helps to understand the PXP model, which is a \emph{partial cube}, i.e.,  a hypercube where all the vertices violating the constraint have been removed -- see Fig.~B7. The resulting PXP graph contains two smaller hypercubes of dimension $L/2$, with $|\mathbb{Z}_2\rangle$ and $|\mathbb{Z}_2'\rangle$ states as extremal vertices, and the polarised state $|{\circ}{\circ}\ldots\rangle$ at the intersection point of the two hypercubes. Additionally, there are also ``bridges" that  connect the two hypercubes, e.g., in Fig.~B7 one such configuration is ${\circ}{\bullet}{\circ}{\circ}{\bullet}{\circ}$.

								\begin{minipage}[c]{0.98\linewidth}
	\begin{center}
		\includegraphics[width=0.4\columnwidth]{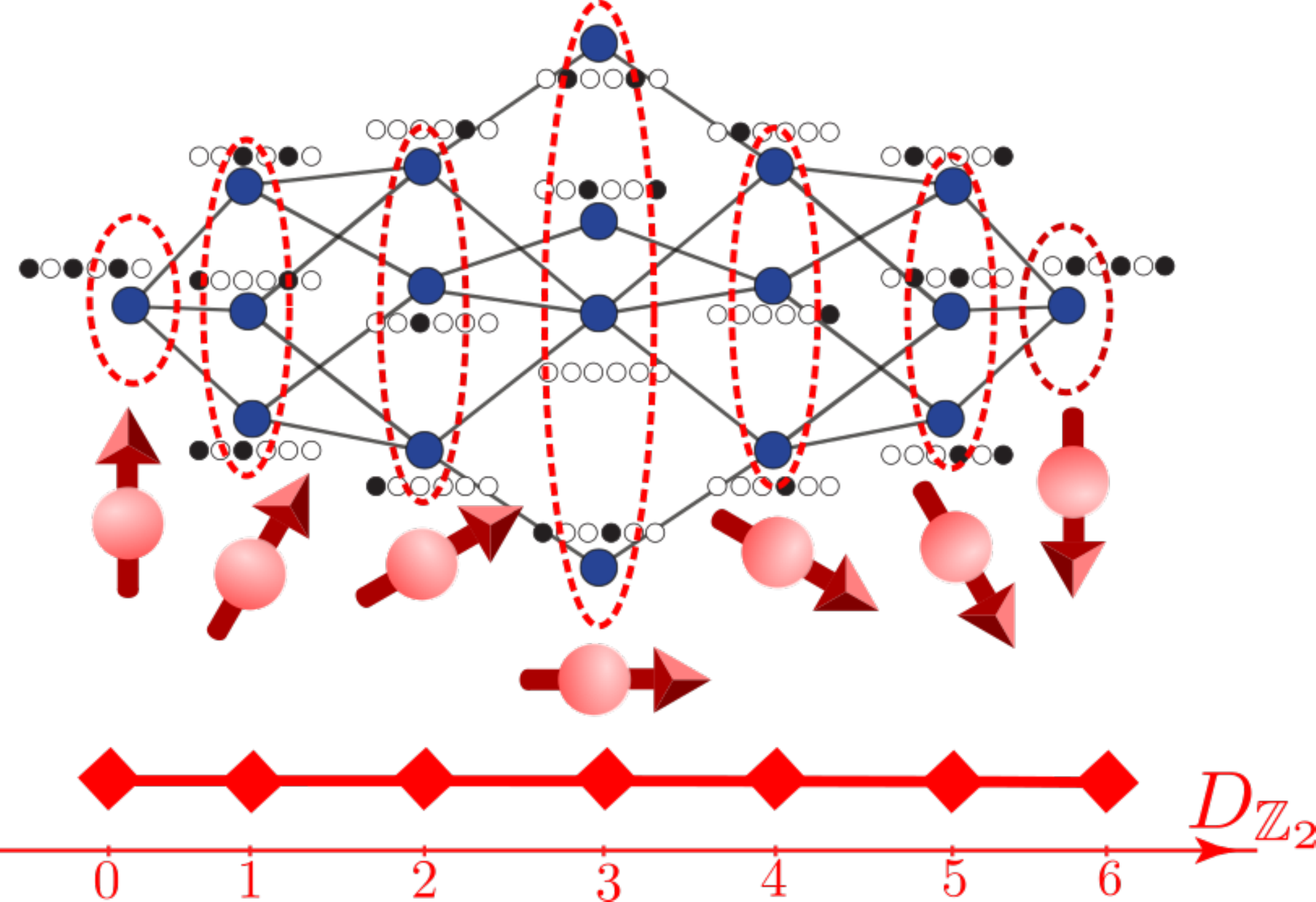}\label{fig:fsa}
	\end{center}
	\vspace*{-0.2cm}
	{\small  
		{\bf Figure B7:} Graph representation of the PXP Hamiltonian for $L{=}6$ spins. Vertices represent all states compatible with the constraint, while lines connect states related by a single spin flip. The FSA models the non-thermal eigenstates and revivals from $|\mathbb{Z}_2\rangle$ state by compressing the graph to a one-dimensional ladder of states, which are identified with basis states of a large spin
		of magnitude $s=\frac{L}{2}$ (red arrows). The states are labelled by their $z$-projection of spin or, equivalently, by their Hamming distance $D_{\mathbb{Z}_2}$ from the leftmost vertex. 
	}
	
\end{minipage}
\vspace*{0.2cm}

\noindent Quantum walk on the PXP graph starting in the $|\mathbb{Z}_2\rangle$ vertex can be accurately modelled by assuming the wave function spreads only in the forward direction, a scheme known as the Forward Scattering Approximation (FSA)~\cite{Turner2017}. The FSA compresses the partial cube down to a one-dimensional chain, where  each site $|n\rangle$ is a superposition of states with the same number of excitations relative to $|\mathbb{Z}_2\rangle$ vertex. Thus, for $L$ atoms, the chain contains $L{+}1$ sites. The FSA interprets each site $|n\rangle$  as an eigenstate of a big spin with magnitude $s{=}L/2$ and pointing at an angle $\pi n/L$ with respect to $z$-axis. As mentioned in the text, the spin-raising and lowering operators directly follow from a decomposition of  the PXP Hamiltonian~\cite{Turner2017}. The FSA not only allows to construct remarkably accurate approximations to tthe exact eigenstates of the PXP and several other models~\cite{Turner2017,TurnerPRB,Moudgalya2019,Bull2019}, but it also provides the foundation for the elegant interpretation of the revivals as precession of an emergent spin.

\end{shaded}

\section{Semiclassical dynamics}\label{sec:semiclassical}

Quantum-classical correspondence has been at the heart of quantum mechanics since the early 20th century. Parallels between the behaviour of a quantum system and its classical counterpart have been extensively studied in the context of few-body systems, such as the free propagating particle in a bounded domain (stadium billiard being a particular example), coupled harmonic oscillators, and a rotor subject to periodic driving~\cite{GutzwillerBook}. In all these cases, the semiclassical limit can be taken by sending $\hbar\to 0$, and the behaviour of the quantum system is often found to closely parallel its classical counterpart. For instance, quantum counterparts of classically chaotic systems can be typically described by random matrix theory~\cite{Mehta2004}.

Intrinsically many-body systems, such as the PXP  chain in Eq.~(\ref{Eq:PXP}), present challenges for the semiclassical description based on the conventional large-$N$ limit or the mean-field approximation. The latter, for example,  ignores any correlations between different spatial degrees of freedom, and thus violates the constraint in the PXP model in Eq.~(\ref{Eq:PXP}). The constraint introduces non-local correlations: if a given spin is in $|{\bullet}\rangle$ state, its neighbours must be in $|{\circ}\rangle$ state. Such correlations build in quantum entanglement, which cannot be ignored unless one relaxes the constraint. 

The MPS methods discussed in Sec.~\ref{sec:mps} allow to incorporate entanglement in a systematic way that goes beyond mean field~\cite{Haegeman}.  For the PXP model, it has been first shown by Ho \emph{et al.}~\cite{wenwei18TDVPscar} that this approach can be tailored to capture dynamics from the initial $|\mathbb{Z}_2\rangle$ state. This state is a period-2 density wave, which can be described using a 2-atom unit cell.  Hence, the variational MPS state is parametrised by two sets of angles, ${\bf x}_i {=} (\theta_i, \phi_i)$, $i{=}1,2$, that describe the state of spins on odd and even sites in the lattice. The MPS matrices can be taken to be~\cite{Michailidis2020}
\begin{equation}\label{Eq:A-matrix}
	{A}^\bullet(\theta_{i}, \phi_{i}) = 
	\left(\begin{matrix}
		0 &  i e^{-i\phi_{i}}\\
		0 & 0
	\end{matrix} \right),
	\
	{ A}^\circ(\theta_{i}, \phi_{i}) = 
	\left(\begin{matrix}
		\cos\theta_{i}  &  0\\
		\sin\theta_{i}& 0
	\end{matrix} \right),
\end{equation}
where we notice that  $A^\bullet {\propto} \sigma^+$ satisfies the condition $A^\bullet (\theta, \phi)A^\bullet (\theta', \phi')=0$, which effectively imposes the constraint that no two adjacent spins are in ``$\bullet$" state. The special points  $(\theta_1,\theta_2)=(0,\pi/2)$ and $(\theta_1,\theta_2)=(\pi/2,0)$ correspond to $|\mathbb{Z}_2\rangle = |{\bullet}{\circ}{\bullet}{\circ}\ldots\rangle$ product state and its  partner shifted by one lattice site, $|{\mathbb{Z}}_2'\rangle = |{\circ}{\bullet}{\circ}{\bullet}\ldots\rangle$.  

By projecting quantum dynamics onto the MPS manifold spanned by Eq.~(\ref{Eq:A-matrix}) using TDVP, as outlined in Sec.~\ref{sec:tdvp}, one obtains a classical non-linear system for $\theta_i(t)$, $\phi_i(t)$. The derivation is cumbersome and the final equations are also rather complicated (many useful technical details can be found in the Supplementary Material of Ref.~\onlinecite{wenwei18TDVPscar}). To give an idea, we quote the final equations of motion from Ref.~\onlinecite{Michailidis2020}
\begin{subequations}\label{Eq:Z2-dyn}
	\begin{eqnarray}\label{Eq:Z2-dyn-a}
		\dot\theta_1 &=& \tan \theta_2 \sin \theta _1 \cos ^2\theta _1 \cos\phi _1+\cos \theta _2 \cos \phi _2, \\ \label{Eq:Z2-dyn-b}
		\dot\phi_1 &=&-\mu_z+ 2\tan \theta _1 \cos \theta_2 \sin \phi _2 -\frac{1}{2} \tan \theta _2 \cos \theta _1
		\left(2\sin^{-2} \theta _2 +\cos 2 \theta _1-5\right)\sin \phi_1,
	\end{eqnarray}
\end{subequations}
which were derived for the PXP model in the presence of a chemical potential,  $\mu_z \sum_j n_j$, where $n_j \equiv \frac{1}{2}(1+\sigma_j^z)$.  The equations for $\theta_2,\phi_2$ can be obtained by substitution $1\leftrightarrow2$.

 Before discussing the solutions of the equations of motion~(\ref{Eq:Z2-dyn}), we note that the expectation value of the Hamiltonian is a conserved quantity, provided the the equations of motion are satisfied. This conservation law effectively reduces the dimensionality of the phase space from four down to three dimensions, as it restricts the  dynamics to constant energy surfaces.  However, when $\mu_z{=}0$ the system in Eqs.~(\ref{Eq:Z2-dyn}) actually has a class of solutions with \emph{two} angles being stationary, $\phi_{1,2}{=}0$~\cite{wenwei18TDVPscar}. This class of solutions corresponds to a flow-invariant subspace, here arising due to particle-hole symmetry and time-reversal invariance of the  PXP Hamiltonian in Eq.~(\ref{Eq:PXP})~\cite{Michailidis2020}. 
 
\begin{figure}[thb]
	\includegraphics*[width=0.7\linewidth]{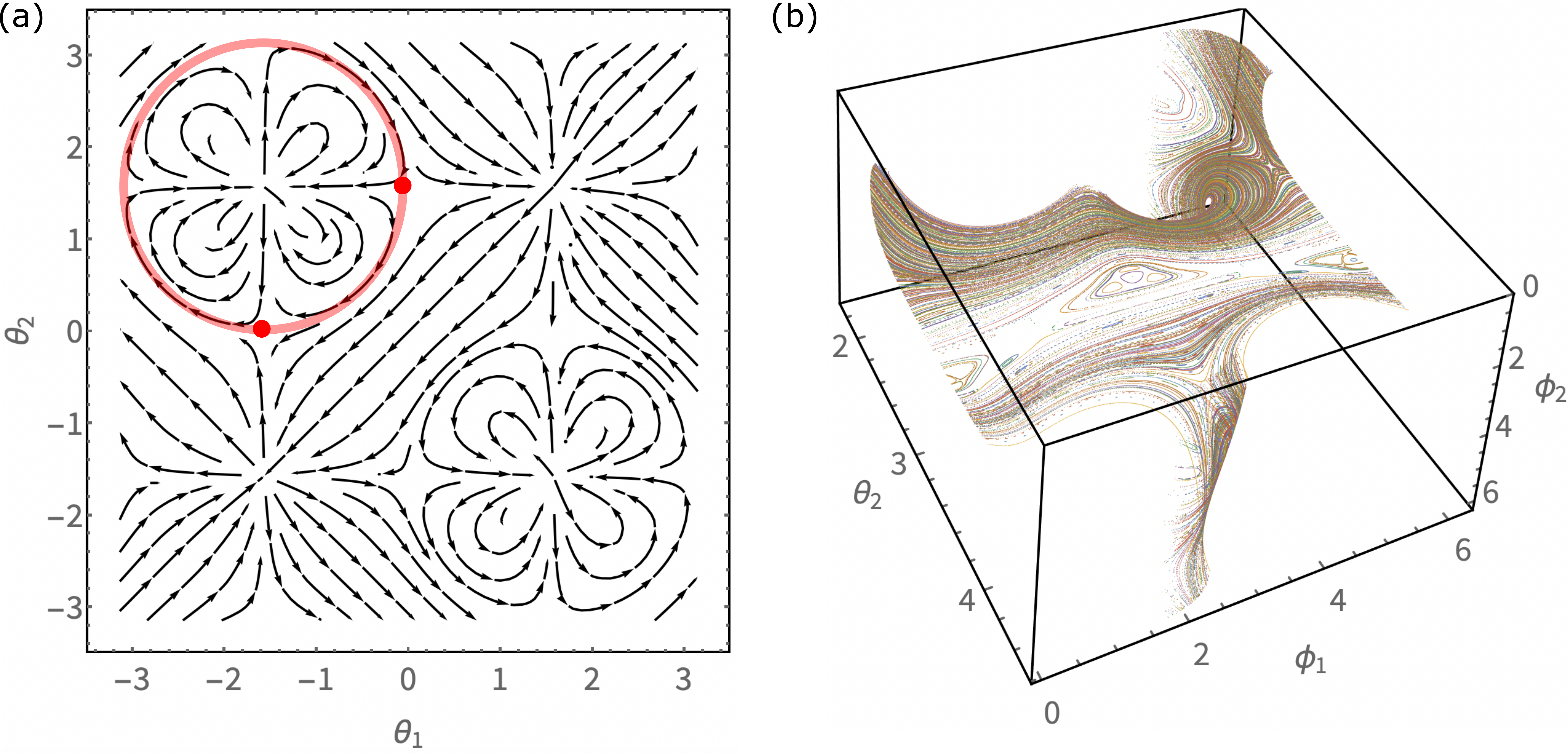}
	\caption{
		(a) Flow diagram obtained from the variational ansatz with two degrees of freedom reveals an unstable periodic trajectory (red line), which is responsible for the fidelity revivals from the period-2 density wave state in the quantum case~\cite{wenwei18TDVPscar}. 
		The periodic trajectory passes through the N\'eel states, $\ket{\mathbb{Z}_2}$ and $\ket{\mathbb{Z}'}$, shown by red dots, and avoids the polarised state, $\ket{0}\equiv |{\circ}{\circ}{\circ}\ldots\rangle$, located at $(0,0)$. 
		As shown in Ref.~\onlinecite{wenwei18TDVPscar}, quantum leakage outside the variational manifold is minimised in the vicinity of the periodic orbit, thus explaining the relevance of this trajectory in the fully quantum dynamics. 
		(b) In the presence of the chemical potential $\mu_z {\neq} 0$, the dynamical system develops mixed phase space containing large regular regions, as shown in Ref.~\onlinecite{Michailidis2020}. The plot shows the associated Poincar\'e section for  $\mu_z {=} 0.325$ at fixed $(\theta^*_1{=}1.25\pi,\dot \theta_1<0)$. 
	}\label{fig:semiclassical}
\end{figure}

 Restricting to the flow-invariant subspace $\phi_{i}{=}0$ (with $\mu_z{=}0$), we plot the flow diagram in $\theta_1$-$\theta_2$ plane in Fig.~\ref{fig:semiclassical}(a). Shown in red colour is the periodic trajectory first identified in Ref.~\onlinecite{wenwei18TDVPscar}. This trajectory is intimately linked with quantum revivals for the initial Ne\'el state $\ket{\mathbb{Z}_2}$ in Fig.~\ref{fig:pxp}(a), since this state  is represented by the point $(\theta_1,\theta_2) {=} (\pi/2,0)$ on the trajectory. We note that the variational description in Fig.~\ref{fig:semiclassical}(a) is not accurate at long times:  the fidelity in the full quantum evolution in Fig.~\ref{fig:pxp}(a) shows a visible decay, which is in contrast with the variational description that implies perfect oscillations. As we mentioned in Sec.~\ref{sec:tdvp}, this discrepancy between the exact quantum evolution and its projection onto the variational manifold can be quantified via quantum leakage in Eq.~(\ref{Eq:leak2}). The leakage measures the instantaneous disagreement between exact dynamics and its projection onto the variational manifold. Remarkably, the periodic trajectory is located entirely in the low leakage region~\cite{wenwei18TDVPscar}, thus providing \emph{a posteriori} justification for the coherent quantum dynamics when the system is prepared in $|\mathbb{Z}_2\rangle$ state.
 
 Two-dimensional phase space, however, is special as it precludes chaos. Dynamical systems with phase spaces higher than  two-dimensional are known to display both classical chaos as well as regions of stable motion --- the so-called Kolmogorov-Arnold-Moser (KAM) tori~\cite{ArnoldBook}. These features are prominent in Fig.~\ref{fig:semiclassical}(b) which shows the Poincar\'e section of the dynamical system in Eq.~(\ref{Eq:Z2-dyn}) when the chemical potential is present ($\mu_z{=}0.325$) and the variables $\phi_i$ cannot be fixed to zero.  To obtain such a graph, we follow the standard recipe: we choose some value for  $\theta_1$  and follow the successive intersections of the flow with the chosen section. This produces a discrete mapping known as the Poincar\'e map -- this maps a given point $(\theta^*_1,\phi_1,\theta_2,\phi_2)$ on the chosen hyperplane to a position  $(\theta^*_1,\phi'_1,\theta'_2,\phi'_2)$ where the trajectory intersects this plane again. Periodic trajectories correspond to stationary points of the Poincar\'e map, while a chaotic system returns to the same plane at a location that is generally far away from the previous encounter. This procedure for evaluating the Poincar\'e sections can be conveniently implemented, e.g., in Mathematica using the built-in \texttt{NDSolve} routine.

 Fig.~\ref{fig:semiclassical}(b) illustrates a typical example of \emph{mixed phase space}~\cite{GutzwillerBook}: there exist (in general, multiple) KAM tori with stable periodic orbits in their centers, surrounded by the chaotic sea. Energy conservation results in complicated surfaces in the space $(\phi_1,\theta_2,\phi_2)$ that cannot be globally projected. Therefore, the figure shows a small part of the Poincar\'e section in the vicinity of the stationary point,  $(\phi_1,\theta_2)\approx(2.92,2.98)$, corresponding to a stable periodic trajectory~\cite{Michailidis2020}.  This periodic trajectory is surrounded by circle-shaped contours which are the intersections of KAM tori by our section plane. Unlike the pure PXP model ($\mu_z{=}0$) where the periodic orbit passes through a product state ($\ket{\mathbb{Z}_2}$), in this case the states on the trajectory are entangled in general (though entanglement is necessarily small, since the orbit passes through MPS states of low bond dimension).

\subsection{Discussion: benefits and pitfalls of TDVP}

To summarise, the TDVP discussed above allows to obtain classical equations of motion and analyse parallels between the dynamics of a quantum system and its projection onto the variational manifold. The presence of (stable or unstable) classical periodic trajectories with small quantum leakage leads to slow relaxation in the full many-body dynamics.

There are several useful insights we glean from TDVP. (i) The KAM tori inform us about special features of quantum dynamics in the underlying model. For example, when dynamics is initialised precisely  on the periodic orbit, there is slower relaxation of local observables and slower entanglement growth in quench dynamics~\cite{Michailidis2020}. Furthermore, initialising the dynamics in the centre of a big KAM torus results in slower entanglement growth. This highlights the role of quantum leakage in determining which of the trajectories give rise to slow thermalisation in exact quantum dynamics.  (ii) Revivals in the PXP model are robust to various deformations of the Hamiltonian, provided that  in the course of the deformation one follows the periodic trajectory that is influenced by the deformation~\cite{Michailidis2020}. (iii)  The TDVP approach has also been fruitful in obtaining a more complete understanding of the PXP dynamics in manifolds with more degrees of freedom. In particular, this approach can describe the dynamics from the period-3 density wave state, $|\mathbb{Z}_3\rangle$, which also shows revivals of fidelity, as seen in Fig.~\ref{fig:pxp}(a). The corresponding variational ansatz for this state includes three angles $\theta_{i}$, $i{=}1,2,3$, which parametrize the state of three atoms within a unit cell (the corresponding $\phi_i$ once again can be fixed to zero in the flow-invariant subspace). While the equations of motion are considerably more complicated than ones in Eq.~(\ref{Eq:Z2-dyn}), their numerical solution also generally gives rise to mixed phase space, similar to Fig.~\ref{fig:semiclassical}(b).

At the same time, one must beware some caveats to the TDVP approach. First, the variational manifold  is not unique --- one can always enlarge the MPS ansatz by including additional parameters that build in  longer-range entanglement. This leads to potential ambiguities in the definition of the trajectory. While some TDVP ans\"atze may find unstable periodic  trajectories, more generic embeddings may result in mixed phase space.   Second, at present there is no understanding under which conditions the quantum leakage is small. Hence, even if one recovers a large number of periodic trajectories, they may all have strong leakage and not lead to any revivals in quantum dynamics, e.g. as in the chaotic Ising model in Ref.~\onlinecite{Michailidis2020}.  

Finally, many questions remain about quantum-classical correspondence in TDVP as a function of bond dimension $\chi$. Complementing the study of weak ergodicity breaking at small $\chi$, TDVP has been used to study thermalisation at larger $\chi$. In particular, Ref.~\onlinecite{Leviatan2017} demonstrated that physical properties saturate in the case of transverse field Ising model when  bond dimension $\chi{>}2$. In addition, Ref.~\onlinecite{Hallam2019} studied the spectrum of Lyapunov exponents in the case of high bond dimension TDVP. The work in Ref.~\onlinecite{Michailidis2020} shows that in the case of low bond dimensions one may encounter non-chaotic behaviour in the TDVP dynamics, limiting the utility of the Lyapunov exponent. On the one hand, one may expect that  mixed phase space does not persist for large bond dimensions, as increasing  $\chi$ increases the dimension of the phase space where the TDVP dynamics occurs, making it more susceptible to chaos. On the other hand, for local Hamiltonians the Lieb-Robinson bound~\cite{LiebRobinson} suggests that a small bond dimension is sufficient to capture quantum dynamics at early times. Thus, it is important to understand how the mixed phase space evolves upon including additional MPS parameters that describe longer-range entanglement.

\section{Quantum many-body scars}\label{sec:scars}

In Sec.~\ref{sec:mechanisms} we discussed different mechanisms that lead to the emergence of non-thermal eigenstates in highly-excited energy spectra of non-integrable systems. In the past few years, the name \emph{quantum many-body scars} has been widely adopted as an umbrella term for such non-thermal eigenstates, regardless of the underlying mechanism of ergodicity breaking. In this section, we explain in more detail the reasons why this name was adopted by Turner \emph{et al.}~\cite{Turner2017}. The inspiration was the PXP model and its many parallels with a single particle confined to a stadium-shaped billiard. In the latter case, the particle's eigenfuctions can concentrate around certain periodic orbits in the limit $\hbar\to 0$, a phenomenon  Heller~\cite{Heller84} called quantum scarring in the early 1980s.  In the remainder of this section, we discuss the semiclassical aspects of the PXP physics that make it the many-body analogue of quantum scarring.

	\subsection{Scars in few-body systems}
	
		What can we learn about the behaviour of a  quantum system by looking at its classical counterpart? The  Bohr-Sommerfeld quantisation demonstrates that classical integrable systems, such as the harmonic oscillator or hydrogen atom, have rather special quantum spectra and eigenstates.  For classically chaotic systems, e.g., the Bunimovich stadium, traditional quantisation methods do not work. Typically, quantum counterparts of classically chaotic systems display level repulsion in their spectrum and their eigenstates look random.  However, short unstable periodic orbits may leave a strong imprint on the system's quantum dynamics and eigenstate properties  -- this influence of classical periodic orbits is known as ``quantum scarring"~\cite{Heller84}: 
	\vspace*{0.1cm}
	
		 \noindent {\bf Definition} (from Ref.~\onlinecite{HellerLesHouches}): A quantum eigenstate of a classically chaotic system has a  `scar' of a periodic orbit if its density on the classical invariant manifolds near the periodic orbit  is enhanced over the statistically expected density.
	\vspace*{0.1cm}
		 
     Utilising this definition, one can detect scars by visualising the wave function probability density, which can be seen to concentrate around certain periodic classical orbits, e.g., the diamond-shaped ``bow tie" orbit in the Bunimovich stadium~\cite{HellerLesHouches}. Moreover, scars leave an imprint on the dynamics: when a wave packet is launched in the vicinity of an unstable periodic orbit, it will tend to cluster around the orbit at later times, displaying a larger return probability than a wave packet launched elsewhere in the phase space. Furthermore, such a wave packet can be expanded over a small number of eigenstates that have approximately similar energy spacing, in contrast to an arbitrary wave packet. Much of this phenomenology was originally acquired through numerical solutions of quantum billliard problems, since the latter are non-integrable and do not admit analytical solution. 
	
	Nevertheless, rigorous proofs~\cite{Schnirelman1974,DeVerdiere1985,Zelditch1987} have established that individual eigenstates of the billiard are ``almost always" ergodic, since the phase space area affected by scarring vanishes in the semiclassical limit $\hbar\to 0$, shrinking around the periodic orbit. These proofs of scarring have been based on constructions of  \emph{approximate} eigenstates or ``quasimodes",  i.e., special states designed to be strongly localised around a classical periodic orbit. For example, in the Bunimovich stadium, some of the quasimodes have the form $\psi(x,y)= \phi(x)\sin(n\pi y)$, i.e., they represent a standing wave in one direction with a suitably chosen envelope function in the other direction. By carefully controlling the density of states, it was possible to show that there exist eigenstates with an anomalously high overlap with a small number of quasimodes, hence ``inheriting" their scarring properties~\cite{HellerLesHouches}.
	
	Although the above suggests that scars are ``rare" anomalies, they do have physical signifance.  For example, scars provide an important counterexample to the intuitive expectation that every eigenstate of a classically chaotic system should locally look like a random superposition of plane waves~\cite{Berry1983}. Furthermore, also counterintuitively, a scarred quantum system appears more ``regular" than its classical counterpart, since in the latter case there is no enhancement of density along the periodic orbit in the long-time limit.  Finally, scars play a role in many experiments, including microwave cavities~\cite{Sridhar1991}, semiconductor quantum wells~\cite{Wilkinson1996}, and the hydrogen atom in a magnetic field~\cite{Wintgen1989}.

\subsection{Quasimodes in the PXP model}

In few body systems, the understanding of quantum scars rests on two pillars: the existence of unstable classical periodic orbits and the quasimodes which leave a scar upon some of the eigenfunctions.  The main difficulty in passing to the many-body case is how to give precise meaning to this terminology: the physics now takes place in the many-particle Hilbert space, precluding ``easy" visualisation of the classical trajectory.   In Secs.~\ref{sec:tdvp} and \ref{sec:semiclassical} we introduced one way of defining the classical trajectory via TDVP, wherein many-body dynamics is projected onto a restricted space of MPS states.  This approach systematically builds in entanglement, controlled by the MPS bond dimension, thus it forms a natural framework to define the semiclassical limit of the PXP model which does have an obvious $\hbar\to 0$ limit. 

The application of the TDVP method to the PXP model successfully captures the revivals following quenches from specific product states, hence it is natural to suggest that this approach defines an effective ``semiclassical'' description of the quantum dynamics, and holds the same relation with the exact PXP model as expected from the Bohr correspondence principle. However, in order to complete the analogy with single-particle scars, it is necessary to show that this limit also gives rise to quasimodes which, in turn, accurately approximate the quantum eigenstates of the PXP model. 

\begin{figure}[tbh]
	\includegraphics[width=0.98\columnwidth]{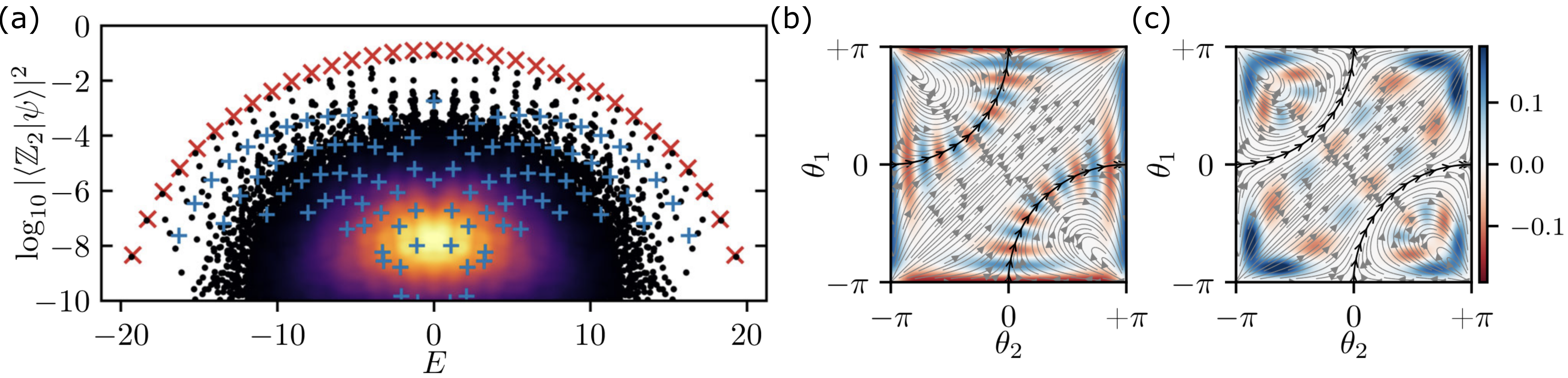}
	\caption{
		(a)  Scatter plot showing energies of all eigenstates of the PXP model vs. their overlap with $|\mathbb{Z}_2\rangle$ state.  Red crosses denote top-band quasimodes, blue pluses are the remaining quasimodes. Both of types of quasimodes are obtained by projecting the PXP model into the symmetric subspace $\mathcal{K}$ defined in the text. Colour indicates the density of data points.
		(b)-(c):     Viewing a selected top-band (b) and non top-band (c) quasimode as wavefunctions over $\theta_1$ and $\theta_2$ for $L{=}128$. Colour scale represents the real part of the wavefunction. Top-band quasimodes concentrate around the classical periodic orbit (black line), displaying quantum scarring. 
		The other quasimodes avoid this periodic trajectory and concentrate around the corners of the square. 
		Note that the $|\mathbb{Z}_2\rangle$ and its translated partner correspond to $(\theta_1,\theta_2){=}(\pi,0)$ and $(0,\pi)$ in the notation of this figure.  Due to the MPS parametrisation, the wavefunction  in (b) visibly spreads out towards the boundary and becomes completely delocalised at the edges, $\theta_i = \pm \pi$, where the points are indistinguishable in the transverse direction.
		Panels reproduced from Ref.~\onlinecite{Turner2020} under a Creative Commons licence CC BY 4.0.
	}\label{fig:quasimodes}
\end{figure}

In Ref.\onlinecite{Turner2020} this correspondence was achieved by constructing a subspace  $\mathcal{K}$, which is fully symmetric over permutations within each of the two sublattices, comprising even and odd sites in the PXP chain -- see Box~8. This forms a mean-field approximation for the PXP model wherein one describes the dynamics in terms of numbers of excitations on even or odd sites, while discarding any detailed local information (such as the precise positions of excitations within a sublattice). This approximation reduces the exponential complexity of the many-body problem down to polynomial in system size, and the quasimodes are obtained by simply diagonalising the Hamiltonian projected into the symmetric subspace, $\mathcal{K} H_\mathrm{PXP}\mathcal{K}$. As can be seen in Fig.~\ref{fig:quasimodes}(a), these quasimodes are excellent approximations to the exact eigenstates of the PXP model, in particular they capture the $L{+}1$ non-thermal eigenstates with the largest projection on $\ket{\mathbb{Z}_2}$ state  in Fig.~\ref{fig:pxp}(a). 

It turns out that the symmetric subspace $\mathcal{K}$ has an intimate relationship with the subspace $\mathcal{M}$ of MPS states discussed in Sec.~\ref{sec:semiclassical}, which can be viewed as a Gutzwiller projection of spin coherent states with a unit cell of two sites~\cite{wenwei18TDVPscar}.  The states in $\mathcal{M}$ are parametrised by two angles which represent probabilities for a site of each sublattice to hold an excitation, while  $\mathcal{K}$ consists of states with definite occupations numbers for each sublattice. In this sense, there is a direct relationship between the two subspaces, reminiscent of the relationship between canonical and grand canonical ensembles in statistical mechanics. More formally, it can be proven that the linear span of $\mathcal{M}$ is equal to the symmetric subspace $\mathcal{K}$, for every fixed system size $L$~\cite{Turner2020}.  This implies that the dynamics within $\mathcal{K}$ necessarily includes all quantum fluctuations on top of the equations of motion in Eqs.~(\ref{Eq:Z2-dyn}).

The correspondence between the classical system $\mathcal{M}$ and the quantum system in its linear span ($\mathcal{K}$) finallly allows us to view the quasimodes $\ket{\psi}\in \mathcal{K}$  as wavefunctions $\psi(\theta,\phi)$:
\begin{eqnarray}
	\psi(\theta,\phi) = \bra{\Psi(\theta,\phi)} S_\mu^{-1/2} \ket{\psi}. \label{eq:frame}
\end{eqnarray}
Here $\Psi(\theta,\phi) \in \mathcal{M}$ is the MPS state parametrised by Eq.~(\ref{Eq:A-matrix}), and $S_\mu$ is the frame operator~\cite{Turner2020}.  This operator is needed because we need to equip the classical state space with an integration measure $\mu$ in order to have a well-defined quantisation. In the more familiar manifolds of quantum states, such as spin coherent states~\cite{Gazeau:2009vb} or unconstrained MPS~\cite{Green2016}, the measure $\mu$ is simply given by the Haar measure for a transitive group action. However, for the subspace $\mathcal{K}$ with angles $\theta_i$, $\phi_i$ having 2-site periodicity, the measure is non-trivial and one must use the frame operator.

Using the above correspondence Eq.~(\ref{eq:frame}), the real part of the quasimode wavefunctions is plotted in Fig.~\ref{fig:quasimodes}(b)-(c). For one of the top-band quasimodes near the middle of the spectrum [denoted by a cross in Fig.~\ref{fig:quasimodes}(a)], we observe enhanced concentration around the $|\mathbb{Z}_2\rangle$  classical trajectory [shown by the black line in Fig.~\ref{fig:quasimodes}(b)].  This behaviour is strikingly reminiscent of wavefunction scarring in quantum billiards. In particular, 
 following the classical trajectory, the phase of the wavefunction winds, and integrality of this winding number leads to the quasimodes' approximately equal spacing in energy. We recognise this behaviour from quantisation of regular trajectories~\cite{GutzwillerBook}: in the old quantum theory, this underpins Sommerfeld-Wilson quantisation and leads to the de~Broglie standing-wave condition for the Bohr model. These features are inherited by the exact scarred eigenstates given their strong overlap with the quasimodes.
   In contrast, Fig.~\ref{fig:quasimodes}(c) shows one of quasimodes deeper in the bulk of the spectrum. Such quasimodes are typically found around the orbits connecting corners of the manifold, where quantum leakage is large (see Sec.~\ref{sec:semiclassical}). It is  expected that these quasimodes would not survive the ``injection'' into the full Hilbert space, i.e., while the top-band quasimodes accurately approximate the true eigenstates of the full PXP model, this is not the case for the other quasimodes.

\definecolor{shadecolor}{rgb}{0.8,0.8,0.8}
\begin{shaded}
	\noindent{\bf Box 8 $|$ Mean field approximation for the PXP model}
\end{shaded}
\vspace{-9mm}
\definecolor{shadecolor}{rgb}{0.9,0.9,0.9}
\begin{shaded}
	\noindent The essence of mean-field theory is the erasure of  unnecessary local information. For example, to describe a free paramagnet, which is obtained by dropping the projectors in the PXP Hamiltonian in Eq.~(\ref{Eq:PXP}),  we only need to know the \emph{total number} of excitations relative to some reference state, not their precise positions in the lattice. This idea has been generalised to describe the reviving dynamics in the PXP model where the important information is the number of excitations on each of the two \emph{sublattices}, comprising even and odd sites in the chain~\cite{Turner2020}. Such an approach is reminiscent of the symmetric subspaces in studies of fully-connected models~\cite{Sciolla2011, Mori2017}. However, the key difference is that in the PXP model the permutation symmetry is broken to the sublattice level, and many of the permutation shuffles violate the constraint and therefore have to be excluded, which makes the analysis more involved.

	To illustrate the approach, the symmetric subspace $\mathcal{K}$ is defined by forming a set of equivalence classes $(n_1, n_2)$, where integers $n_1$, $n_2$ label the number of excitations on the two sublattices, encompassing the odd and even sites, respectively. Elements in these classes are equivalent under the action of the product of two symmetric groups $S_{L/2}$ which ``shuffle" the sites in each sublattice. For example, states $\ket{{\bullet}{\circ}{\bullet}{\circ}{\circ}{\bullet}{\circ}{\circ}{\circ}{\circ}}$ and $\ket{{\circ}{\circ}{\bullet}{\circ}{\circ}{\circ}{\bullet}{\circ}{\circ}{\bullet}}$ belong to the same class as they both have two excitations in the first sublattice and one in the second. 
	An example of the construction of $\mathcal{K}$ for the PXP model of size $L{=}8$ is presented in Fig.~B8. The orthonormal basis for  $\mathcal{K}$ is  built from symmetric combinations of members of each class, $\ket{(n_1,n_2)} \propto \sum_{x\in(n_1,n_2)}\ket{x}$, where the sum runs over all product states $\ket{x}$ that are members of the class $(n_1,n_2)$.  Class sizes and matrix elements of the PXP Hamiltonian projected to $\mathcal{K}$ can be calculated using combinatorics~\cite{Turner2020}, allowing for a highly-efficient description of the PXP model and its dynamical properties.
	
	\vspace*{0.1cm}

									\begin{minipage}[c]{0.98\linewidth}
		\begin{center}
			\includegraphics*[width=0.5\linewidth]{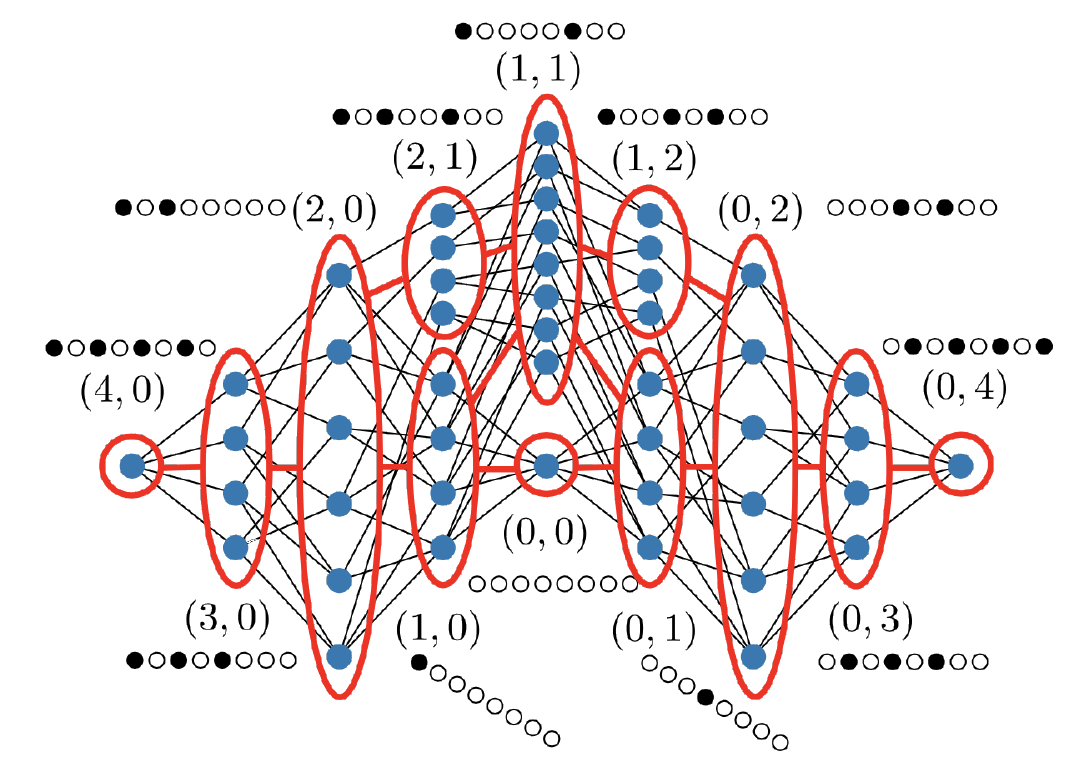}
		\end{center}
		\vspace*{-0.2cm}
		{\small {\bf Figure B8:} Construction of the symmetric subspace $\mathcal{K}$ in the PXP model for $L{=}8$ spins. Each blue dot represents an allowed product state compatible with the constraint. The states are grouped into equivalence classes, denoted by ellipses. Each class is labelled by a representative state, e.g., the class $(0,3)$ contains the representative state ${\circ}{\bullet}{\circ}{\bullet}{\circ}{\bullet}{\circ}{\circ}$ and all other obtained by permuting sites in each of the two sublattices. Reproduced from Ref.~\onlinecite{Turner2020} under a Creative Commons licence CC BY 4.0. }
	\end{minipage}
	\vspace*{0.2cm}

\end{shaded}

\subsection{Discussion:  scars or not? }

The results presented in this section and Sec.~\ref{sec:semiclassical} provide justification for referring to the  non-thermal eigenstates in the PXP model as a many-body analogue of single particle scars. One of the crucial ingredients that allows to make such a connection is the quantum-classical correspondence. At the same time, in Sec.~\ref{sec:mechanisms} we introduced a number of other models that also feature non-thermal eigenstates, quite reminiscent of those in the PXP model. Here we briefly discuss the main similarities and differences of these models.  

The PXP model, with its generalisations to higher spins~\cite{wenwei18TDVPscar} and clock models~\cite{Bull2019},  is broadly similar to models whose non-thermal eigenstates form an SGA. The main difference is that the SGA in the PXP model is  only \emph{approximate}, and the subspace has weak residual couplings to the thermal bulk. In addition, the PXP model also realises an \emph{approximate} Krylov subspace built upon $|\mathbb{Z}_2\rangle$ root state.  The Krylov subspace becomes exact when generated by $H^+$ in Eq.~(\ref{eq:hplus}). Finally, a single PXP eigenstate at zero energy in the middle of the spectrum can be viewed as projector-embedded AKLT ground state~\cite{Shiraishi_2019}. While similar connections between scarring mechanisms may be anticipated in other models mentioned in Sec.~\ref{sec:mechanisms}, their demonstration remains an open problem. 

In particular, the existence of classical periodic trajectories underlying quantum revivals in the PXP model calls for exploration of semiclassical dynamics in other scarred models discussed in Sec.~\ref{sec:mechanisms}.  For example, in the spin-1 XY model~\cite{Iadecola2019_2} in Eq.~(\ref{Eq:1XY}), the initial state with perfect revivals was identified to be 
\begin{equation}
	\ket{\psi_0} = \bigotimes_j \frac{1}{\sqrt{2}} \left( \ket{1}_j - (-1)^j  \ket{-1}_j \right) ~.
\end{equation}
This state is a superposition of the scar tower states $\ket{\mathcal{S}_n^{XY}}$ in Eq.~(\ref{eq:ISXYscar1}) and it can be prepared as the ground state of the Hamiltonian $H = Q^\dagger {+} Q \propto J^x$. Since these states are equally spaced in energy with spacing $2h$ [see Eq.~(\ref{Eq:1XY})], time evolving $\ket{\psi_0}$ gives perfect revivals with frequency $2h$. For the AKLT tower of scars, such ``rotations" in the space of scar states do not seem to produce a very simple initial state. However, as shown in Ref.~\onlinecite{MotrunichTowers} the AKLT scar tower  can be compressed into a state with finite MPS bond dimension (e.g., bond dimension equal to 8 in a system with periodic boundary conditions). 

We expect that similar constructions of reviving initial states can be performed for other models with an SGA using the properties of $Q^\dagger$.  By contrast, for non-thermal eigenstates produced via the projector-based embedding scheme, we generally do not expect underlying periodic trajectories. Similarly, models with Krylov fragmentation (Sec.~\ref{sec:krylov}) are not guaranteed to have revivals, even if the root state $|\psi_0\rangle$ is experimentally preparable. This is because general tridiagonal matrices do not support revivals, unless their matrix elements are tuned to special values~\cite{Christandl2004}. Thus, the study of classical periodic trajectories in a broader family of models could be helpful as a finer classification scheme for models displaying weak ETH violation in their eigenstate properties, in particular as a way of distinguishing many-body scarring from more generic embeddings of non-thermal eigenstates.

\section{Weak ergodicity breaking in experiment}\label{sec:exp}

The experiments that opened the door to the investigation of weak ergodicity breaking discussed above were performed using quantum simulators based on Rydberg atom arrays~\cite{Labuhn2016, Bernien2017}. Here we provide a brief overview of these experiments, focusing on the work by Bernien \emph{et al.}~\cite{Bernien2017} which reported the first experimental observation of quantum many-body scarring. Following this, we discuss more recent experiments realising the tilted Fermi-Hubbard in optical lattices, where signatures of non-ergodicity due to Hilbert space fragmentation have been detected~\cite{Scherg2020}. 

\subsection{Rydberg atoms}

\begin{figure}[t]
	\includegraphics[width=0.5\columnwidth]{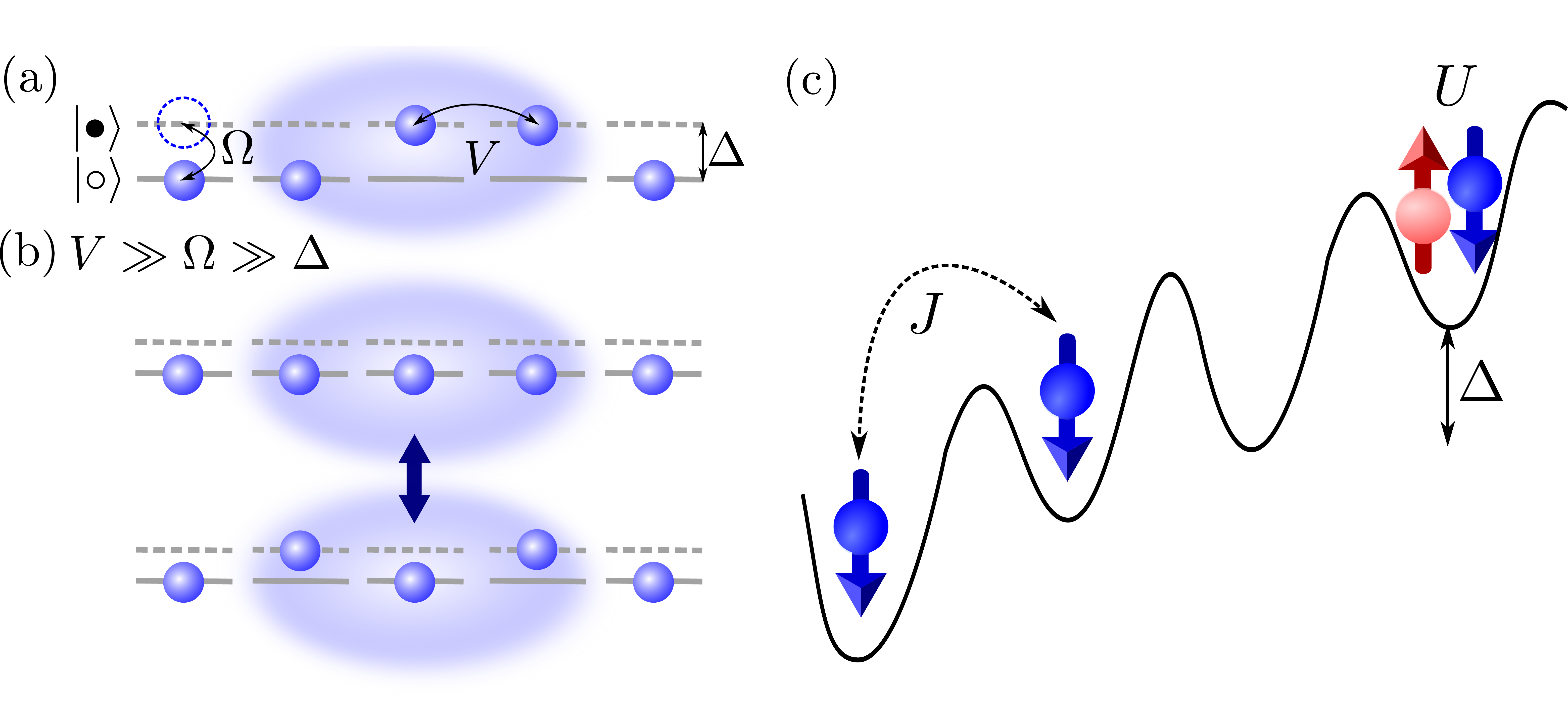}
	\caption{
		(a) Rydberg atom as a two level system, where the empty circle corresponds to the ground state and the filled circle denotes the excited Rydberg state. The Rabi frequency is denoted as $\Omega$ and the detuning of the $|\bullet\rangle$ state from resonance is $\Delta$. The Rydberg blockade (shaded) leads to a shift of the energy level where both atoms are in the $|\bullet\rangle$-state, bringing it strongly off-resonance. (b) When the Rydberg blockade is strong ($V\gg \Omega \gg \Delta$), excitations of neighbouring atoms are energetically suppressed. In this limit, the effective model describing the Rydberg atom chain is the PXP model in Eq.~(\ref{Eq:PXP}). (c) Fermi-Hubbard model realised in an optical lattice~\cite{Scherg2020}, with nearest neighbour hopping $J$ and on-site interaction $U$. Tilt of the lattice is controlled by parameter $\Delta$ [with a different meaning from the detuning $\Delta$ in (a)-(b) and Eq.~(\ref{eq:rydberg})!]. 
	} 
	\label{fig:exp} 
\end{figure}

An individual Rydberg atom may be viewed as an effective two-level system, where the two states  $|{\circ}\rangle$,  $|{\bullet}\rangle$ correspond, respectively, to an atom in the ground state and an atom in the so-called Rydberg state, i.e., with a highly excited electron in the outer shell, see Fig.~\ref{fig:exp}(a). The state of an individual Rydberg atom may be manipulated via coupling to circularly-polarised radiation which causes Rabi oscillations with frequency $\Omega$. The relative detuning $\Delta$ of the Rydberg state off-resonance can also be controlled,  leading to an effective single atom Rydberg Hamiltonian $H = (\Omega/2) \sigma^x +\Delta \hat n$, where the operator $\sigma^x = |{\circ}\rangle\langle{\bullet}|+|{\bullet}\rangle\langle{\circ}|$ is the Pauli matrix which describes Rabi oscillations,  and $\hat n = |{\bullet}\rangle\langle{\bullet}|$ measures the population of the Rydberg state. 

Atoms in Rydberg states interact via dipole-dipole interactions, thus the many-body Hamiltonian describing this system is given by 
\begin{eqnarray}\label{eq:rydberg}
H = \frac{\Omega}{2} \sum_i \sigma^x_i - \Delta \sum_i \hat{n}_i + \sum_{i<j} V_{ij} \hat{n}_i \hat{n}_j,
\end{eqnarray}
where $V_{i,j}=1/r_{ij}^6$ and $r_{ij}$ is the distance between the atoms. By tuning $r_{ij}$ one can achieve the regime of the Rydberg blockade, as illustrated in Fig.~\ref{fig:exp}(a). In this regime, the shift of an energy level with two excited Rydberg atoms is so strong that the state $|{\bullet}{\bullet}\rangle$ is off-resonant and cannot be reached from the ground state. Rydberg blockades with varying radii were demonstrated experimentally~\cite{Labuhn2016, Browaeys2020} via the frequency renormalization of the Rabi oscillations. Coherent dynamics was observed in small arrays of Rydberg atoms in both 1D and 2D when the blockade radius exceeded the linear size of the chain.

More recent experiments~\cite{Bernien2017}  have studied Rydberg arrays in a qualitatively different regime with a relatively \emph{short} radius of the Rydberg blockade. In particular, long Rydberg chains, containing up to $51$ atoms with the nearest-neighbour blockade, were quenched from a period-2 density wave initial state, $|\mathbb{Z}_2 \rangle \equiv |{\bullet}{\circ}{\bullet}{\circ}\ldots\rangle$, corresponding to the state with the maximal possible number of Rydberg excitations allowed by the blockade.  The dynamics of the domain wall density, where domain walls are defined as adjacent 
$|{\circ}{\circ}\rangle$ or $|{\bullet}{\bullet}\rangle$ configurations (the latter is excluded in the regime of perfect blockade), revealed long-time oscillations, similar to Fig.~\ref{fig:pxp}(a). 

The density revivals observed in Ref.~\onlinecite{Bernien2017} were challenging to explain theoretically, for the following reasons. First, the density wave state $|\mathbb{Z}_2\rangle$ forms an infinite-temperature ensemble for the atoms in the Rydberg blockade, thus it should display fast equilibration. Indeed, for other initial states at infinite temperature, such as $|{0}\rangle \equiv |{\circ}{\circ}{\circ}\ldots\rangle$, the domain wall density was found to relax  quickly to the steady-state value. Second, the persistence of crystalline order far from equilibrium was surprising because the Rydberg atom system was not known to have any conserved quantities other than the total energy. Finally, the frequency of domain wall density oscillations did not coincide with $\sqrt{2}\Omega$, which would be na\"{i}vely  expected for collective oscillations of two atoms in the Rydberg blockade regime, signalling that many-body effects played an important role.  As we have already seen, these puzzles are now understood based on the studies of the PXP model summarised in Sec.~\ref{sec:pxp}. The PXP model arises as the effective model for the Hamiltonian in Eq.~(\ref{eq:rydberg}) when $V_{i,i}$ is infinite and interactions beyond nearest-neighbour are set to zero~\cite{Lesanovsky2012}.  An important  step towards the explanation of the  observed oscillations was made by identifying the non-thermalising eigenstates in the spectrum of the PXP model in Fig.~\ref{fig:pxp}(b). The core phenomenology of the PXP model -- the small number of ETH-violating eigenstates within the thermalising spectrum and the presence of many-body revivals and slow relaxation in quenches from specific initial states -- led to the identification of the phenomenon in Ref.~\onlinecite{Bernien2017} with a many-body analogue of  quantum scarring~\cite{Heller84}. 

We note that since the experiment in Ref.~\onlinecite{Bernien2017} there have been further advances in similar Rydberg platforms. In Ref.~\onlinecite{Dolev2020} it was shown that coherent revivals associated with quantum many-body scars can be enhanced via periodic driving, which generates a robust subharmonic response akin to discrete time-crystalline order~\cite{ElseReview}. The driving is realised by modulation of the detuning term, $\Delta(t) = \Delta_0 + \Delta_m \cos(\omega_m t)$, and the enhancement is found in the non-perturbative regime of $\Delta_m, \Delta_0, \omega_m \sim \Omega$. It was found that such time-dependent detuning can lead to a five-fold increase of scar lifetime compared to the fixed-detuning case~\cite{Dolev2020}.  Notably, parametric driving not only delays thermalisation, but also alters the actual trajectory
being stabilised, which can be used to effectively ``steer" complex dynamics in many-body systems~\cite{Maskara2021}. This opens the door to robust creation and control of complex entangled states in the exponentially large Hilbert spaces of many-body systems, with intriguing potential applications in areas such as quantum metrology. Other fundamental phenomena recently realised in the same Rydberg platform include dynamical phase transitions and Kibble-Zurek scaling~\cite{Keesling2019} and topological spin liquids~\cite{Semeghini2021} (see also Ref.~\onlinecite{Satzinger2021} for a realisation in superconducting qubits.)

\subsection{Tilted optical lattices}

Another platform for probing weak ergodicity breaking  are ultracold atoms in tilted optical lattices, which realise the 1D Fermi-Hubbard model~\cite{Scherg2020}:
\begin{eqnarray}\label{eq:hamfull}
H =  -J \sum_{j,\, \sigma=\uparrow ,\, \downarrow} \hat{c}^\dagger_{j,\sigma}\hat{c}_{j+1,\sigma}^{} +{\rm h.c.}+\Delta \sum_{j, \sigma} j\hat{n}_{j,\sigma}   
	+  U\sum_j{\hat{n}_{j,\uparrow} \hat{n}_{j,\downarrow} }, \;\;\;\;\;
\end{eqnarray}
where $\hat{c}_{j,\sigma}^\dagger$ denotes the usual electron creation operator on site $j$ with spin projection $\sigma$, $\hat n_{j,\sigma} \equiv \hat{c}_{j,\sigma}^\dagger \hat{c}_{j,\sigma}$ , $J$ and $U$ are the hopping and on-site interaction terms, respectively. Compared to the standard Hubbard model, Eq.~(\ref{eq:hamfull}) includes the tilt potential $\Delta$, which has the form of a dipole term, $\sim j \hat  n_{j}$. 

The experimental setup in Ref.~\onlinecite{Scherg2020} consists of a degenerate Fermi gas of $^{40}\mathrm{K}$ atoms that is prepared in an equal mixture of two spin components. The atoms are loaded into a 3D optical lattice from which 1D chains are isolated  along $x$-direction with the length of about 290 lattice sites.  The on-site interaction strength $U$ is controlled by a Feshbach resonance and a magnetic field gradient is used to create the tilt $\Delta$ (approximately independent of spin). We note, however, that linear potentials similar to the tilt one in Eq.~(\ref{eq:hamfull}) can also be realised in other platforms such as trapped ions~\cite{Morong2021} and superconducting qubits~\cite{Guo2020}, where signatures of strong ergodicity breaking have been recently observed and attributed to ``Stark many-body localisation"~\cite{StarkMBL1,StarkMBL2}.

The model in Eq.~(\ref{eq:hamfull}) is a natural setting for exploring Hilbert space fragmentation. Ignoring the spin degree of freedom, in the limit of large tilt, $\Delta \gg U, J$, the leading-order off-diagonal term~\cite{MoudgalyaKrylov} is precisely the pair-hopping Hamiltonian in Eq.~(\ref{eq:pairhopping}).  With spin included, the effective Hamiltonian in the limit of large tilt (at third order) comprises an off-diagonal term~\cite{Scherg2020}
\begin{align} \label{H3exp}
	&\hat T_3= \sum_{i,\sigma}{ \hat c_{i,\sigma} \hat c_{i+1,\sigma} ^{\dagger} \hat c_{i+1,\bar{\sigma}} ^{\dagger} \hat c_{i+2,\bar{\sigma}}} +\text{h.c.},
\end{align}
whose strength is proportional to $J^{(3)}=J^2U/\Delta^2$ ($\bar \sigma$ denotes the opposite spin of $\sigma$). At the same order, the effective Hamiltonian also contains another off-diagonal term, $\propto 2 J^{(3)} \hat T_{XY}$ where $\hat T_{XY} =  \sum_{i,\sigma} \hat c_{i,\bar{\sigma}}^{\dagger} \hat c_{i+1,\bar{\sigma}} \hat c_{i+1,\sigma}^{\dagger}  \hat c_{i,{\sigma}}$, as well as two diagonal terms, 
$U\Big( 1 -\frac{4 J^2}{\Delta^2}\Big)\sum_{i} \hat n_{i,\uparrow} \hat n_{i,\downarrow}$ and  $2 J^{(3)} \sum_{i,\sigma} \hat n_{i,\sigma} \hat n_{i+1,\bar{\sigma}}$. 

The effective model in Eq.~(\ref{H3exp}) conserves the dipole moment, $\sum_{i, \sigma} i\hat{n}_{i, \sigma}$. The fact that the hopping rate $J^{(3)}$ is proportional to the interaction strength highlights that interactions are necessary to generate dipole-conserving processes~\cite{Scherg2020}.  For initial states that are products of singlons, the connected dynamical sector $\mathcal{S}$ only represents a vanishing fraction of the whole (effective) symmetry sector, thus severely restricting the dynamics of the system. The dipole-conserving processes in general involve the generation of doublons. This is, however, penalised by the Fermi-Hubbard on-site interaction and therefore, we expect a slowing down of the dipole-conserving dynamics.  The additional spin-exchange $\hat T_{XY}$ increases the connectivity, but cannot fully connect the whole dipole symmetry sector and the system remains fragmented.

In experiment, ergodicity breaking was probed by quenching the system from an initial state  chosen to be a charge-density wave of singlons on even sites (filling factor $\nu{=}1/2$), which is prepared using a bichromatic optical superlattice. The initial state can be described as an incoherent mixture of site-localised particles with random spin configuration. The subsequent evolution is monitored by extracting the spin-resolved imbalance $\mathcal{I}^{\sigma}=(N_{e}^{\sigma}-N_{o}^{\sigma})/N^\sigma$, where $N_{e(o)}^{\sigma}$ denotes the total number of spin-$\sigma$ atoms on even (odd)  sites and $N^\sigma=N_{e}^{\sigma}+N_{o}^{\sigma}$. 
A non-zero steady-state imbalance signals a memory of the initial state, where $\mathcal{I}^\sigma(t{=}0)=1$.  In the regime  $\Delta\gg J, |U|$, the effective Hamiltonian is dipole-conserving up to third order in $J/\Delta$, which was argued~\cite{Scherg2020} to be responsible for the non-ergodic behaviour of imbalance $\mathcal{I}$, which did not decay to its thermal value at moderate times.  At much later times, higher-order processes, beyond the ones in Eq.~(\ref{H3exp}), are expected to reconnect the different Krylov sectors. 

Tilted Fermi-Hubbard model is a promising platform to investigate a range of ergodicity breaking beyond fragmentation.
Although experimentally challenging due to finite evolution times, it would be interesting to reconcile the phenomenon of Stark MBL and Hilbert-space fragmentation, by studying the impact of weak disorder or residual harmonic confinement on the long-time dynamics.  Moreover, it would be interesting to explore the connection between lattice gauge theories and Hilbert-space fragmentation, which could be addressed experimentally in a similar system~\cite{Yang2020QLM}. Finally, recent work~\cite{Desaules2021} has proposed that quantum many-body scars could also be realised in the same model, albeit at a different filling factor compared to Ref.~\cite{Scherg2020} -- see Box~9.

\newpage

\definecolor{shadecolor}{rgb}{0.8,0.8,0.8}
\begin{shaded}
	\noindent{\bf Box 9 $|$ Proposal for scars in the Fermi-Hubbard model }
\end{shaded}
\vspace{-9mm}
\definecolor{shadecolor}{rgb}{0.9,0.9,0.9}
\begin{shaded}
	
\newcommand{\HG}{\ket{{-}{+}}}
\newcommand{\HGG}{\ket{{+}{-}}}
\newcommand{\isol}{\ket{\downarrow 2 \uparrow}}
\newcommand{\isoll}{\ket{\uparrow 2 \downarrow}}

\noindent	In the regime $\Delta{\approx} U{\gg} J$, the tilted Fermi-Hubbard model in Eq.~(\ref{eq:hamfull}) was argued to host many-body scars at filling factor $\nu{=}1$~\cite{Desaules2021}.	In this case the sum of the dipole moment and the number of doublons is effectively conserved, giving rise to the effective Hamiltonian
	\begin{eqnarray}\label{eq:hameff}
	H_{\rm eff} &=&-J\sum_{j,\sigma}  \hat{c}^{\dagger}_{j,\sigma}\hat{c}_{j+1,\sigma}\hat{n}_{j,\overline{\sigma}}(1-\hat{n}_{j+1,\overline{\sigma}}) + \mathrm{h.c.},
	\end{eqnarray}
	where hopping to the left (which decreases the total dipole moment by 1) is only allowed if it increases the number of doublons by the same amount ($\bar{\sigma}$ is opossite spin from $\sigma$). 
	
	The action of the Hamiltonian in Eq.~(\ref{eq:hameff}) within the $\nu{=}1$ sector fragments the Hilbert space beyond the simple conservation of $U{+}\Delta$. The largest connected component of the Hilbert space is the one containing the state with alternating $\uparrow$ and $\downarrow$ fermions. In such large sectors, after resolving all the symmetries, the model can be shown to be non-integrable~\cite{Desaules2021}. 

	The proposal for many-body scars in the effective Hamiltonian in Eq.~(\ref{eq:hameff}) is based on the existence of a regular subgraph which has the form of the hypergrid -- a Cartesian product of line graphs (in our case, of length 3), i.e., the hypergrid is isomorphic to an adjacency graph of a free spin-1 paramagnet, as illustrated below.	
	
						\begin{minipage}[c]{0.98\linewidth}
		\begin{center}
			\includegraphics*[width=0.4\linewidth]{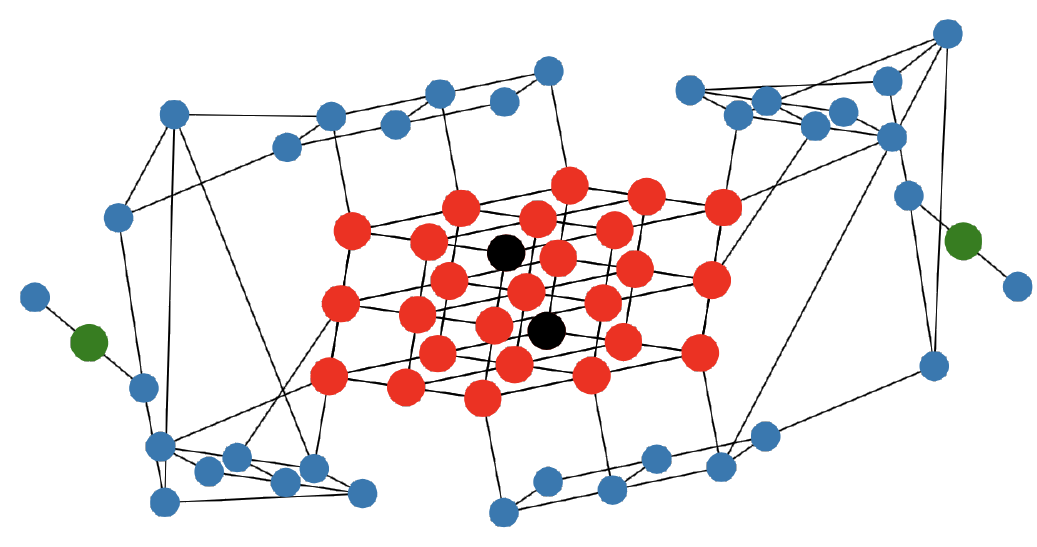}	
		\end{center}
		{\small {\bf Figure B9:} Adjacency graph of the model in Eq.~(\ref{eq:hameff}) for $L{=}6$. Red vertices denote the states belonging to the hypergrid, with the black vertices corresponding to $\HG$, $\HGG$ states defined in the text. Green vertices are the isolated states $\isol$, $\isoll$ which live on the tails of the graph. 
		}
	\end{minipage}
\vspace*{0.2cm}

	The mapping to the hypergrid follows from analysing pairs of sites, which can be in states  $``-"\equiv  ({\downarrow}{ \uparrow}) $, $``2" \equiv ({\updownarrow}{0}) $ or $``+"\equiv ({\uparrow}{ \downarrow})$, leading to a three level system. 	 Note that the configuration $({0}{\updownarrow})$ is omitted, as doublons can only be formed by hopping to the left. Inside the hypergrid, there are two states for which the cell alternates between $-$ and $+$. These are the state $\HG\equiv \ket{{-}{+}{-}{+}\ldots}=\ket{\downarrow \uparrow \uparrow \downarrow\downarrow \uparrow \uparrow \downarrow\ldots}$ and its spin-inverted partner, $\HGG=\ket{{+}{-}{+}{-}\ldots}\equiv \ket{ \uparrow\downarrow \downarrow \uparrow \uparrow\downarrow \downarrow \uparrow\ldots}$.  The states $\HG$ and $\HGG$ are shown in black colour in figure below. These two states are the only corners of the hypergrid (state with only $+$ and $-$ cells) with no edges going out of it. When the system is initialised in either of these states, it was numerically shown to exhibit persistent oscillations in the quench dynamics and other scarring phenomenology similar to the PXP model discussed in Sec.~\ref{sec:pxp}.

\end{shaded}

\section{Conclusions}\label{sec:conclusions}

The goal of this chapter was to present a pedagogical introduction to the new kind of dynamical behaviour in many-body systems -- weak ergodicity breaking. While we refrained from giving a formal definition, weak ergodicity breaking was intuitively introduced as a strong dependence of relaxation dynamics on the system's initial configuration. We contrasted this behaviour against conventional ergodic systems, by showing that certain many-body states can be long-lived and they exhibit parametrically slow relaxation, unlike  other initial configurations that quickly reach thermal equilibrium. Many recent studies discussed above, theoretical as well as experimental, have revealed that a number of familiar physical systems do not conform to the expectations of the ETH in its strong form, in that such systems can host non-thermal eigenstates and exhibit long-time coherent dynamics.  The common pattern that arises in these diverse physical systems is an emergent non-thermalising subspace, which is dynamically decoupled from the rest of the many-body Hilbert space.  Quantum entanglement, in particular, has been the vital tool for identifying such non-thermal subspaces in different models. 

At this stage, a complete classification of non-thermal subspaces and their underlying mechanisms is still lacking.  In particular, more work is needed to understand the connection between many-body scarring and two broad classes of weak ergodicity breaking phenomena: theories with confinement~\cite{Rahul17,Calabrese16, Konik2, Yang2020, CastroAlvaredo2020} and lattice gauge theories~\cite{Magnifico2020, Chanda2020, Borla2020}.  The latter, in particular the one-dimensional quantum link model -- which has recently been realised in a Bose-Hubbard quantum simulator~\cite{Yang2020QLM} -- have intriguing connections with the PXP model~\cite{Surace2019}. It would  be interesting to explore possible connections with PXP  in higher dimensions~\cite{Celi2020}, as well as to understand the relation with kinetic constraints, which can generally lead to slow glassy-like dynamics~\cite{Pancotti2019,Roy2020,
	Lan2017_2,Feldmeier2019, Hart2020}.

On the theory side, many questions remain about the quantum-classical correspondence in the physical systems discussed above. 
The TDVP variational approach complements recent efforts in understanding parallels between classical chaos measures, on the one hand, and thermalising quantum dynamics and its underlying transport coefficients on the other hand~\cite{Leviatan2017,Hallam2019}.  While many-body scarred models are quantum chaotic, their properties deviate from other chaotic models such as the Sachdev-Ye-Kitaev (SYK) model~\cite{Sachdev1993,Kitaev2015}, which has attracted much attention as the fastest scrambler of quantum information~\cite{Maldacena2016}. Such deviations could be identified and studied more systematically using the variational approach and its resulting mixed phase portraits, which appear to be a generic feature of local Hamiltonians.
One of the outstanding challenges is to bring together this approach to real-time dynamics with approaches that target the many-body eigenstates, possibly using  quantum entanglement and other quantum information techniques as a way of  linking quantum revivals and eigenstate properties~\cite{Alhambra2019}. 

Finally, there is strong experimental and practical interest in weak ergodicity breaking. For example, many-body scarred revivals provide a mechanism for maintaining coherence, despite the presence of interactions which normally scramble local quantum information. In particular, scars in Rydberg chains  have already been utilised for the preparation of specific entangled states~\cite{Omran570}. This application made use of quantum control based on the variational TDVP approach and its identification of entangled periodic trajectories that simultaneously have small quantum leakage. Thus, scars may have a wider range of applications, for example in protected state transfer on quantum networks or in quantum sensing~\cite{Dooley2021}. Such applications require deeper theoretical understanding of the effects that protect the coherence of scars, as well as the development of  general experimental techniques for creating them on demand, e.g., using periodic driving in Rydberg arrays~\cite{Dolev2020}, pumping protocols in dipolar Bose gases~\cite{Kao2020} or by tilting the Fermi-Hubbard model~\cite{Scherg2020}.

\vspace*{0.2cm}

\section{Acknowledgements}

I would like to thank my collaborators Kieran Bull, Soonwon Choi, Jean-Yves Desaules, Wen Wei Ho, Ana Hudomal, Mikhail Lukin, Ivar Martin,  Hannes Pichler, Nicolas Regnault, Ivana Vasi\'c,  and in particular Dmitry Abanin, Alexios Michailidis, Maksym Serbyn and Christopher Turner. Moreover, I am grateful to Maksym Serbyn for designing Fig.~\ref{Fig1}, and to Alexios Michailidis for help with replotting Fig.~\ref{fig:semiclassical}(b).  I would also like to acknowledge useful discussions with B. Andrei Bernevig, Paul Fendley, Thomas Iadecola and Lesik Motrunich.  This work has been supported by the Leverhulme Trust Research Leadership Award RL-2019-015.
This research was supported in part by the National Science Foundation under Grant No. NSF PHY-1748958.

\bibliography{references}

\end{document}